\documentclass[
pre,
twocolumn,
superscriptaddress,
]{revtex4-1}

\usepackage{times}
\usepackage{subcaption}
\usepackage{ragged2e}
\DeclareCaptionJustification{justified}{\justifying}
\usepackage{algorithm}
\usepackage{algpseudocode}
\usepackage{amsmath}
\usepackage{amssymb}
\usepackage{amsthm}
\usepackage{bbm}
\usepackage{enumerate}
\usepackage{enumitem}
\usepackage{pgfplots, pgfplotstable}
\usepackage{placeins}
\usepackage[titletoc,title]{appendix}
\usepackage{wrapfig,lipsum}
\usepackage{graphicx}
\usepackage{dcolumn}
\usepackage{bm}
\usepackage{overpic}
\usepackage{multirow}
\usepackage{tabularx,booktabs}
\usepackage{url}
\usepackage{tabularx}

\usepackage{float}
\floatstyle{plaintop}
\restylefloat{table}

\begin{document}

\title{Effects of setting temperatures in the parallel tempering Monte Carlo algorithm}

\author{Ignacio Rozada, Maliheh Aramon}
\affiliation{1QB Information Technologies (1QBit), Vancouver, BC, V6C 2B5, Canada}

\author{Jonathan Machta}
\affiliation{Physics Department, University of Massachusetts, Amherst, Massachusetts 01003, USA}
\affiliation{Santa Fe Institute, Santa Fe, New Mexico, 87501, USA}

\author{Helmut G. Katzgraber}
\affiliation{Microsoft Quantum, Microsoft, Redmond, Washington, 98052, USA}
\affiliation{Department of Physics and Astronomy, Texas A\&M University, College Station, Texas, 77843-4242, USA}
\affiliation{Santa Fe Institute, 1399 Hyde Park Road, Santa Fe, New Mexico, 87501, USA}

\date{\today}

\begin{abstract} 

Parallel tempering Monte Carlo has proven to be an efficient
method in optimization and sampling applications. Having an optimized
temperature set enhances the efficiency of the algorithm through 
more-frequent replica visits to the temperature limits. The approaches for finding an 
optimal temperature set can be divided into two main categories. The methods of the first 
category distribute the replicas such that the swapping ratio between neighboring replicas is constant and independent 
of the temperature values. The second-category techniques including the feedback-optimized method, on the other hand, 
aim for a temperature distribution that has higher density at simulation bottlenecks,
resulting in temperature-dependent replica-exchange probabilities.  In
this paper, we compare the performance of various temperature setting methods 
on both sparse and fully connected spin-glass problems as well as fully connected 
Wishart problems that have planted solutions. These include two classes of problems that have either
continuous or discontinuous phase transitions in the order parameter.
Our results demonstrate that there is no performance advantage for the methods 
that promote nonuniform swapping probabilities on spin-glass problems where the order parameter has a smooth transition
between phases at the critical temperature. However, on Wishart problems
that have a first-order phase transition at low temperatures, the feedback-optimized method 
exhibits a time-to-solution speedup of at least a factor of two over the other approaches.
\end{abstract}


\maketitle

\section{Introduction}
\label{sec:introduction} 

The parallel tempering (PT) Monte Carlo algorithm---also known as
replica-exchange Monte Carlo \cite{Geyer91}---has been used effectively in solving a
wide range of problems in various fields, such as physics, materials
science, chemistry, logistics, and engineering~\cite{Swendsen86,Hukushima96,Earl05,Katzgraber09e}. 
More recently, it has been repurposed as an effective heuristic for solving hard optimization 
problems ~\cite{moreno2003finding,mandra2016strengths,Zhu16y}. In fact, the
winning entry of the 2016 MaxSAT competition used PT Monte Carlo as
the underlying algorithmic engine of its optimization heuristic. While
the method is extremely powerful, a careful tuning of the parameters is
needed to ensure optimal run times \cite{Rathore05,Hamze10}. In this work, we discuss the effects
of setting the temperature values for the algorithm on paradigmatic spin-glass
problems to elucidate which temperature setting approach is best suited
for a particular problem type.

The PT algorithm simulates $M$ replicas of the original
system at different temperatures, with periodic exchanges based on a 
Metropolis criterion between neighboring temperatures. Replica-exchange 
moves allow replicas to perform a random walk in temperature space, thereby efficiently overcoming energy
barriers. Replicas at high temperatures improve algorithmic mixing,
whereas replicas at low temperatures can reach equilibrium on a shorter time scale compared to 
typical simulations at a fixed low temperature \cite{Kofke04,Predescu05,Fiore08,Machta09,Machta11}. 
The performance of the PT algorithm, like other MC simulation techniques, is highly dependent on its parameters, including the
distribution of replicas in temperature space.  

The approaches for setting the temperature values can be divided into 
two categories. The first category contains techniques that aim to distribute 
the replicas such that the probability of swapping adjacent replicas is independent of the temperature values, is 
equal across all adjacent pairs of replicas, and is sufficiently high so as 
to ensure frequent exchanges. The commonly used geometric distribution 
is considered to be a good approximation for temperature sets if the specific heat of the 
system is roughly constant~\cite{Predescu04}. Other examples include the iterative methods presented by Hukushima and 
Nemoto~\cite{Hukushima96} and Hukushima~\cite{Hukushima99}. In the 
former method, the acceptance ratios are measured using short MC runs, and the 
nearby inverse temperatures are iteratively placed farther or closer if their 
corresponding acceptance ratio is higher or lower than the average 
acceptance ratio, respectively. In the latter approach, referred to as the 
energy method, short MC runs are used to estimate the average energy 
as a function of the inverse temperature for deriving constant probabilities across replicas. 

The second category comprises approaches that discard the premise of 
the first category that an independent and constant swapping ratio
maximizes the extent to which replicas mix. The key idea of the approaches in the second category 
is to increase the performance of the algorithm by minimizing the round-trip
time between the extremal temperatures for each replica. This 
translates to identifying the bottlenecks through measuring the local
diffusivity of the replica ensemble and successively arranging temperature values closer to the
bottlenecks of the simulation~\cite{Trebst04,Katzgraber06}. To our knowledge, there is no study in the literature
that provides insights into the characteristics of problem types that can benefit 
most from the approaches in the second category, particularly when using PT as
an optimization heuristic. 

In this paper, we compare the performance of different approaches 
to determine the distribution of the replicas in the PT algorithm using both sparse 
and fully connected spin-glass problems, as well as fully connected Wishart problems 
that have planted solutions~\cite{Hamze_in_prep}.

The paper is organized as follows. In Sec.~\ref{sec:methods}, we 
provide an overview of temperature setting methods. We introduce our benchmarking 
problems in Sec.~\ref{sec:experimental_setup}, followed by a presentation 
and discussion of the results in Sec.~\ref{sec:results_discussion}. 
Sec.~\ref{sec:conclusion} concludes the paper.

\section{Methods}
\label{sec:methods}
In the PT algorithm, $M$ independent replicas of the system are 
simultaneously simulated at different temperature values $T_1, T_2, \ldots, T_M$. 
After performing a fixed number of MC sweeps at each temperature, swap moves between neighboring 
temperatures, $T_{i-1}$ and $T_i$, are proposed and accepted with probability 
\begin{align}
\mathbbm{P}(E_{i-1}, T_{i-1} \leftrightarrow E_i, T_i) = \min(1, e^{\Delta \beta \Delta E}), & \notag
\end{align}
\noindent where $\Delta \beta = 1/T_i - 1/T_{i-1}$ and $\Delta E = E_i - E_{i-1}$. 

As a result of the replica-exchange moves, replicas with high-energy states tend to 
be moved to high temperatures for diversification and the replicas with low-energy 
states tend to be moved to low temperatures for refinements. Although the simulation 
of $M$ replicas takes $M$ times more CPU time, the PT algorithm 
often has a faster time to solution and better scaling when compared to the simulated annealing algorithm \cite{Wang15}.

In this section, we describe four methods used in the literature to set the temperatures for each replica of the PT algorithm.

\subsection{The feedback-optimized method}
\label{subsec:feedback}

The goal of the feedback-optimized method is to find a temperature set
${T_i}$ such that the average round-trip time between two extremal
temperatures is minimized~\cite{Katzgraber06}. The primary concept of the
feedback-optimized method is similar to that of the adaptive algorithm presented
in the context of ensemble optimization~\cite{Trebst04}, where the local
diffusivity is measured to identify the bottlenecks in a feedback loop, after
which the temperatures are placed near the simulation bottlenecks.

\begin{figure}[h!]
\centering
\includegraphics[width=1\linewidth]{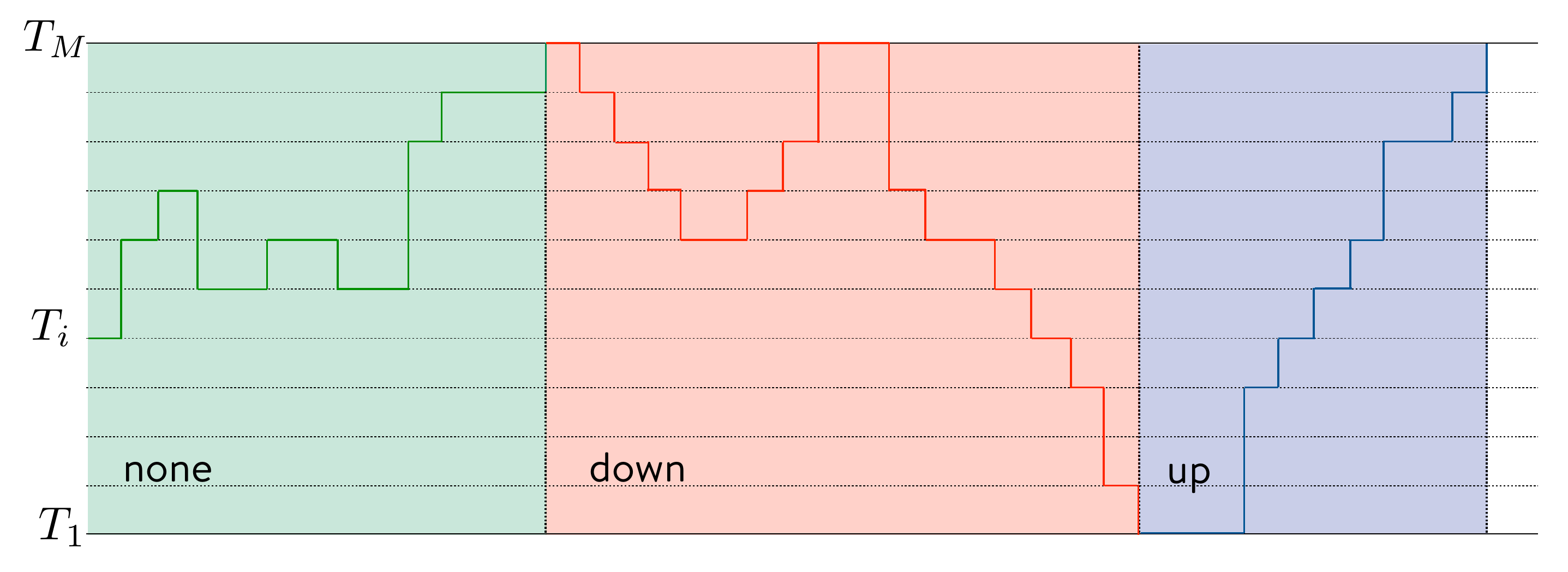}
\caption{An example of replica $i$'s trajectory during the simulation. Its label is
initially set to ``none,'' changes to ``down'' on visiting the highest temperature, and remains unchanged
before hitting the lowest temperature, at which point it is set to ``up.''}
\label{fig:feedback_example}
\end{figure}

To measure the rate at which a replica visits a given temperature, a label of either ``up" or ``down"
is assigned to the replica when it first visits the lowest ($T_1$) or the highest
($T_{\mbox{\scriptsize{M}}}$) temperature. The label of the replica does not change until its first
visit to the opposite temperature extreme. As illustrated in Fig.~\ref{fig:feedback_example}, the
label of replica $i$ is initially set to ``none,'' changes to ``down'' on its first visit to the highest
temperature, remains unchanged until the replica reaches the lowest temperature, and is then
set to ``up.'' We define $n_{\mbox{\scriptsize{up}}}(T_i)$ and $n_{\mbox{\scriptsize{down}}}(T_i)$ as
the number of replicas with labels ``up'' and ``down,'' respectively, that visit temperature $T_i$ during 
the simulation. The fraction of replicas that have recently visited the lowest temperature before 
visiting $T_i$ is given below. We refer to $f(T_i)$ as ``flow'' in this paper, given by
\begin{align}
f(T_i) & = \frac{n_{\mbox{\scriptsize{up}}}(T_i)}{n_{\mbox{\scriptsize{up}}}(T_i) + n_{\mbox{\scriptsize{down}}}(T_i)}  \notag
\end{align}

Following Fick's law of diffusion, which states that the net rate of particles moving through an area 
equals the gradient times the diffusion constant~\cite{Atkins06}, the rate $j$ at which a replica visits a given
temperature $T$ can be stated as
\begin{align}
j &= D(T) \eta(T) \frac{df}{dT},  \label{eq:rate_of_visit} 
\end{align}
\noindent where $D(T)$ is the local diffusivity and $\eta(T)$ is the probability density function for replicas 
visiting temperature $T$. Defining the round-trip time  $\tau = \frac{1}{j}$, we have 
\begin{align}
\tau = \frac{1}{D(T) \eta(T) \frac{df}{dT}}. & \notag 
\end{align}
\noindent Rearranging the above expression yields the following:
\begin{align}
\tau \int_T \frac{df}{dT} dT = \int_T \frac{1}{D(T) \eta(T)} dT. & \notag 
\end{align}
\noindent Since $\int_T \frac{df}{dT} dT = 1$, we have
\begin{align}
\tau = \int_T \frac{1}{D(T) \eta(T)} dT. & \notag
\end{align}

To find a temperature set that minimizes the round-trip time subject to a proper probability
distribution of replicas, the feedback-optimized method solves the following optimization problem:
\begin{align}
& \min_{\substack{T}} \int_T \frac{1}{D(T) \eta(T)} dT& \notag \\
& \mbox{s.t.}   \int_T \eta(T) dT= 1. & \notag 
\end{align}
Defining the Lagrange function as 
\begin{align}
\mathcal{L}(T, \lambda) = \int_T \frac{1}{D(T) \eta(T)} dT - \lambda \big[\int_T \eta(T)dT -1\big], & \notag
\end{align}
\noindent we have 
\begin{align}
\frac{\partial {\mathcal{L}}}{\partial T} = \frac{1}{D(T) \eta(T)} - \lambda \eta(T) = 0, & \notag 
\end{align}
\noindent which results in 
\begin{align}
\eta(T) = \frac{1}{\sqrt{\lambda D(T)}}. & \label{eq:inverse_root_diff}
\end{align}

The solution of the Lagrange function shows that at the optimal temperature set, the optimal density
function $\eta(T)$ is proportional to the inverse square root of the diffusion constant. Following the
assumption that  $\eta(T) = C/\Delta T$ in Ref.~\cite{Katzgraber06} and based on 
Eq.~\eqref{eq:rate_of_visit}, we can further show how to estimate the probability density function by
measuring the flow as
\begin{align}
\frac{1}{D(T)} \propto \frac{\frac{df}{dT}}{\Delta T}\,, 
\end{align}
and substituting it in Eq.~\eqref{eq:inverse_root_diff} gives the following:

\begin{align}
\eta(T) \propto {\sqrt{\frac{\frac{df}{dT}}{\Delta T}}}. & \notag 
\end{align}

It can be also shown that 
\begin{align}
j & = D(T) \eta(T) \frac{df}{dT} & \notag \\
  & = \frac{1}{\lambda \eta^2(T)} \eta(T) \frac{df}{dT} \longrightarrow \eta(T) \propto \frac{df}{dT} \propto \frac{1}{\Delta T}, & \notag
\end{align}
\noindent which implies that at the optimal temperature set, the flow has a constant decay, that is,
$f(T_i) - f(T_{i+1}) = 1/(M-1)$, where $M$ is the number of replicas. We refer to this flow as
the ``optimal flow'' where the flow at replica $i~(i=1, \ldots, M)$ equals $ 1- (i-1)/(M-1)$.

To find the optimal temperature set in practice, we follow the iterative steps presented in 
Algorithm~\ref{alg:feedback} which is based on the algorithm discussed in~Ref.~\cite{Katzgraber06}, but with 
some modifications. The purpose of the algorithm is to find the temperature values 
by first measuring the flow values and then estimating the probability distribution until 
convergence.  

\begin{algorithm}[H]
\scriptsize
\caption{\,\,\scriptsize{Feedback-Optimized Method}}
\label{alg:feedback}
\begin{algorithmic}[1] 
\State {initialize an arbitrary temperature set $\{T_1, \ldots, T_i, \ldots,
T_M\}$ in ascending order and set the number of repeats $n = 1$.} 
\Repeat 
\State perform the PT simulation and measure the flow vector $\{f(T_i)\}$. \label{pt_step_feedback}
\State smooth the flow vector. 
\State estimate the probability density function 
	\begin{align}
		\eta(T_i) = C \sqrt{\frac{f(T_i) - f(T_{i+1})}{T_{i+1} - T_i}}. & \notag 
	\end{align}
\State find $C$ to ensure that
	\begin{align}
		\int_{T_1}^{T_M} \eta(T) dT = 1. & \notag
	\end{align}
\State find the new temperature ${T_i}$ such that 
	\begin{align}
		\int_{T_1}^{T_i} \eta(T) dT = \frac{i-1}{M-1}. & \notag
	\end{align}
\Until{$n = n_{\scriptsize{\mbox{max}}}$}
\end{algorithmic}
\end{algorithm}

To smooth the flow vector, we first filter out the flow value of replica $i$ if it is 
not within $\pm \epsilon~(0.5)$ of the optimal flow value $f^*_i = 1 - (i-1)/(M-1)$. 
More specifically, we remove $f(T_i)$ from the flow vector and $T_i$ from the temperature vector 
if $f(T_i)\notin [f^*_i - \epsilon, f^*_i + \epsilon]$. We then interpolate the flow as a monotonic function 
of temperature using the remaining flow and temperature values based on the method 
presented in~Ref.~\cite{Steffen90}. Denoting the interpolated function as $g$, the 
flow of replica $i$ with temperature $T_i$, $f(T_i)$, is then updated to $g(T_i)$ before 
proceeding to the next step in the algorithm.

Repeating the core procedure of Algorithm~\ref{alg:feedback} $ n_{\scriptsize{\mbox{max}}}$
 times, each with a large number of sweeps, the final temperature vector will be the one that has
the lowest $L^2$-norm distance from the optimal flow vector.

\subsection{The energy method}
\label{subsec:energy}
Hukushima~\cite{Hukushima99} proposed a method that is simpler than the feedback-optimized method 
for determining the temperature values of the parallel tempering replicas. The scheme starts from 
an initial arbitrary temperature distribution, fixes the extremal temperatures, and iteratively adjusts the 
intermediate temperatures such that the replica-exchange probability for all adjacent temperatures is equal. 
Since this scheme is based on estimating the energy at each replica as a function of its 
inverse temperature, we refer to the method as ``the energy method.'' This method falls in the category of 
approaches that aim for a uniform swap rate across replicas. 

Let $\beta_i$ and $E_i$ denote the inverse temperature and the average energy of replica $i$,
respectively. The goal of the energy method is to adjust $\beta_i$ such that the replica-exchange 
probabilities of the adjacent temperatures are equal,
\begin{align}
\mathbbm{P}(E_{i-1}, \beta_{i-1} \leftrightarrow E_i, \beta_i) = \mathbbm{P}(E_i, \beta_i \leftrightarrow E_{i+1}, \beta_{i+1}). & \label{eq:equal_pt_prob} 
\end{align}

More specifically, the replicas are divided into two groups: odd and even. 
Fixing the inverse temperatures of one group, the inverse temperatures of the 
other group are then adjusted in an alternating fashion. The detailed procedure is outlined
in Algorithm~\ref{alg:energy}.

\begin{algorithm}[H]
\scriptsize
\caption{\,\,\scriptsize{Energy Method}}
\label{alg:energy}
\begin{algorithmic}[1] 
\State {initialize an arbitrary inverse temperature set $\{\beta_1, \ldots,
\beta_i, \ldots, \beta_M\}$ in descending order and set the number of repeats $n = 1$.} 
\Repeat 
\State perform the PT simulation and record the energies observed at each replica. \label{pt_step_energy}
\State estimate the average energy as a monotonic function of the inverse temperature. 
\State \label{fix} fixing the odd replicas' inverse temperatures, find the new inverse temperatures ($\beta^{\prime}_i$) for the even replicas using Eq.~\eqref{eq:equal_pt_prob}. 
\State \label{update} update $\beta_i$ of the even replicas to the average of their previous values and $\beta^{\prime}_i$. 
\State perform steps \ref{fix} and \ref{update}, fixing the even replicas' and updating the odd replicas' inverse temperatures. 
\Until{$n = n_{\scriptsize{\mbox{max}}}$}
\end{algorithmic}
\end{algorithm}

The energy scheme is repeated $n_{\scriptsize{\mbox{max}}}$ times, each with a
relatively small number of sweeps. The final temperature values are then determined by taking the
average temperature values of the last $k$ repetitions. 

\subsection{Base methods}
\label{subsec:base}
Two of the base methods for setting the temperatures in PT are the geometric and inverse-linear schemes.

 \begin{itemize} 
	\item Geometric: This scheme is the most commonly used method, where the temperature scheduled is 
	a geometric progression between the low and the high temperatures. Denoting
	 $T_1$ and $T_M$ as the low and the high temperatures, respectively, the intermediate
	 temperatures are set such that $T_i = T_1 R^{i-1}$, where $R = (T_M/T_1)^{1/(M-1)}$.
	\item Inverse linear: In this scheme, the inverse temperatures are placed uniformly between the low and the
	high inverse temperatures, with \mbox{$\beta_i = \beta_ M + (\beta_1 - \beta_M) \times (i-1)/(M-1)$}.
 \end{itemize} 

\section{Benchmarking Problems}
\label{sec:experimental_setup}

We have studied the performance of various temperature setting methods using
synthetic Ising problems, which are considered the simplest hard Boolean optimization
problems. The goal is to find optimal assignments, of $+1$ or $-1$, to decision
variables such that the Hamiltonian (i.e., the cost function) represented below is
minimized. The problem is encoded in a graph $G = (V, E)$, where $V$ is the set of vertices and $E$ is the set of edges. Each decision
variable $s_i$ corresponds to one vertex, and the problem biases $h_i$ and
couplers $J_{ij}$ represent the weights of both the vertices and the edges in the
graph, respectively,

\begin{align}
\mathcal{H} =  - \sum_{(i,j) \in E} J_{ij} s_i s_j - \sum_{i \in V} h_i s_i. & \notag
\end{align}

Finding an optimal variable assignment on a nonplanar graph is an NP-hard
problem~\cite{Pardella08,Liers10,Juenger01}, while exact,
polynomial-time algorithms exist for planar graphs~\cite{Groetschel87}. The PT
algorithm used in this paper solves the Ising problems in their binary
representations, where each decision variable $s_i$ is replaced by its binary
counterpart $x_i$ such that $s_i = 2x_i - 1$. 

Our benchmarking problem set includes five problem categories:

 \begin{enumerate} 

\item[] {\em 3D-bimodal} --- Three-dimensional spin-glass problems on a cubic lattice
with periodic boundaries. The couplings have a bimodal distribution, taking
either a value of $+1$ or $-1$ with equal probability.

\item[] {\em 4D-bimodal} --- Spin-glass problems on a four-dimensional hypercubic 
lattice with periodic boundaries and bimodally distributed couplings.

\item[] {\em SK-bimodal} --- Spin-glass problems on a complete graph, known as
Sherrington--Kirkpatrick (SK) spin-glass problems~\cite{Sherrington75}, where
couplings are chosen from a bimodal distribution. 

\item[] {\em SK-Gaussian} --- SK spin-glass problems, where couplings are
distributed according to a Gaussian distribution with a mean of $0$ and a standard
deviation of $1$, scaled by a factor of $10^5$. 

\item[] {\em SK-Wishart} --- Fully connected Ising problems with planted ground-state 
solutions, and tunable algorithmic hardness~\cite{Hamze_in_prep}. The couplings are generated 
according to a specific type of correlated multivariate Gaussian distribution. The algorithmic hardness 
of this class of problems is determined by a parameter $\alpha$ that specifies the ratio of the 
number of equations to the number of variables. When $\alpha < 1$, the Wishart problem class has 
a first-order temperature transition and exhibits an easy-hard-easy hardness profile. 
We have considered instances with $\alpha = 0.75$ in our benchmarking study.
 \end{enumerate} 

In all problem categories, the biases $h_i$ are chosen to be zero. We have used the 
time-to-solution (TTS) as a performance metric~\cite{ronnow2014defining}
to compare the temperature setting methods. The metric TTS measures the time required
to observe the best-known energy at least once with a probability of $0.99$. Let us 
denote the probability of observing the best-known energy solutions in a run by 
$\theta$ and the time it takes to perform one run by $\tau$. We then have:
\begin{align}
\mbox{TTS} &= \frac{\log(1-0.99)}{\log(1-\theta)} \times \tau. & \notag
\end{align}

In order to estimate the distribution of the success probability $\theta$, we have applied
a Bayesian inference technique to data from a set of sample runs.
A detailed description of how to measure the TTS can be found in~Ref.~\cite{Aramon19}. To
further investigate the relationship between various temperature distributions
and problem sizes, we have considered three problem sizes for all spin-glass problems 
and two for the Wishart problems. For each problem size, $100$ instances have been
randomly generated for the given graph adjacency and disorder. 

Except for the Wishart problems where the ground state energies are known 
by design, the best-known energies for the majority of problem instances have 
been obtained using the PT algorithm with a large number of sweeps ($5 \times 10^5$). 
For small problem instances of the 3D- and 4D-bimodal problem categories, we have considered the optimal energies obtained via 
the spin-glass server~\cite{SG} as the best-known energies.

\section{Results and Discussion}
\label{sec:results_discussion}

In this section, we compare the performance of the PT algorithm on the five problem 
classes where its temperature schedule is set using four different methods: energy, 
feedback optimized, geometric, and inverse linear. Two of the
methods, geometric and inverse-linear (see Sec.~\ref{subsec:base}), generate 
static distributions that  depend only on the values of the extremal temperatures 
and the total number of temperature values. 
The other two methods, feedback-optimized and energy 
(see Sec.~\ref{subsec:feedback} and Sec.~\ref{subsec:energy}, respectively), 
are dynamic algorithms that iteratively adjust the values of the nonextremal
temperatures by performing MC simulations, while keeping the extremal
temperature values and the number of temperature values constant throughout.

\subsection{Experimental setup}
\label{subsec:experimental_setup}

Our experiments consist of two parts: temperature schedule generation and 
performance comparison. More specifically, we have first generated a temperature set for 
each problem instance and for each temperature setting method. We have then 
compared the performance of PT on each problem class using the temperature 
schedules generated by each method.

\vspace{5pt}
\paragraph*{Temperature schedule generation} To generate a temperature schedule for a 
given problem instance, we have used the same temperature bounds and number of temperatures 
in different temperature setting methods. To experimentally find the best values of these parameters, 
we have searched over the parameter space using a geometric schedule and a subset of problem instances. 
The performance metrics including number of 
instances solved to the best-known energy, residual energy, and success probability have been used to select the 
best values for the high and the low temperatures as well as the number of temperatures. 

For the sparse \mbox{3D-bimodal} and 4D-bimodal problem classes, we have tuned the parameters 
per problem size. However, tuning the parameters of dense problem classes per problem size 
is computationally expensive. As a result, for the dense problem instances, that is, SK-bimodal and SK-Gaussian, 
the parameters have been tuned according to the problem class. The temperature parameters 
for all problem classes are shown in Table~\ref{table:temp_parameters}.

\begin{table}
\centering
 \begin{tabularx}{\columnwidth}{XXXXXX}
 \hline \hline
 \multicolumn{2}{X}{Problem Class} 				& Size & $M$ & $T_1$ & $T_M$  \\ \hline
 \multirow{3}{*}{3D-bimodal}					& & 216 	&    10 		&	0.6			&  	2 \\
 										& & 512 	&    10		&	0.6			&      2 \\
										& & 1000 	&    20 		&	0.6			&   	2.5 \\ \hline
\multirow{3}{*}{4D-bimodal}					& & 256 	&    15 		&	0.5			&   	2 \\
 										& & 625 	&    15		&	0.5			&   	2.5 \\
									        & & 1296 	&    30 		&	0.6			&   	3 \\ \hline
 \multirow{3}{*}{SK-bimodal} 				        & & 256 	&    60		&	2			&   	80 \\
 									        & & 576 	&    60 		&	2			&      80 \\
										& & 1024 	&    60 		&	2			&      80 \\ \hline
 \multirow{3}{*}{SK-Gaussian} 					& & 256 	&    60 		& 	$5 \cdot 10^5$	&  $10^7$ \\
 										& & 576 	&    60 		& 	$5 \cdot 10^5$	&  $10^7$ \\
										& & 1024 	&    60 		& 	$5 \cdot 10^5$	&  $10^7$\\ \hline
 \multirow{2}{*}{Wishart} 						& & 64 	     	&   30		&   	0.115 		&   1.4 \\
										& & 128 	&    30		&  	0.115		&   1.4 \\ \hline \hline
  \end{tabularx}
 \caption{High ($T_M$), low ($T_1$), and number of ($M$) temperatures for each problem class and size. 
 The 3D and 4D parameters have been obtained by maximizing the fraction of problems solved to optimality 
 and minimizing the effort needed to find the best solution. To measure the effort for the PT algorithm, 
 we have multiplied the number of sweeps by the number of temperature values. 
 The SK parameters are from~Ref.~\cite{Aramon19} and the Wishart parameters 
 are based on~Ref.~\cite{Hamze_in_prep}. In tuning the parameters of all problem classes and sizes, 
 the temperatures have been distributed based on a geometric schedule.}
 \label{table:temp_parameters}
 \end{table}

The geometric and inverse-linear temperature sets for each problem instance have been directly 
calculated based on the formulas in Sec.~\ref{subsec:base}. As expected, these static methods 
result in equivalent temperature sets for problem instances that have 
the same extremal temperatures and number of temperatures. 
The feedback-optimized temperature set, for each problem instance, has been 
obtained by implementing Algorithm~\ref{alg:feedback}, with 
$n_{\scriptsize{\mbox{max}}}$ set to 5, and the parallel tempering in step~\ref{pt_step_feedback} has 
been performed using $20\,000$ sweeps. The feedback-optimized method occasionally diverges after 
a few iterations when two replicas are placed sufficiently far apart so as to cause the flow to become disconnected.
To alleviate the effect of such a divergence, we have selected the temperature set that has the flow closest 
to the optimal flow as the final temperature schedule. The temperature set generated by the energy method, for each problem instance, 
has been obtained by implementing Algorithm~\ref{alg:energy}, with $n_{\scriptsize{\mbox{max}}}$ set to $500$, 
and with $200$ sweeps of PT in step~\ref{pt_step_energy}. An initial geometric temperature set has been used 
in both the feedback-optimized and energy methods. The maximum numbers of iterations and sweeps for the energy and 
 feedback-optimized methods have been set such that both methods perform the same 
number of sweeps in total in determining the temperature schedule.

\subsection{Temperature schedule convergence}
\label{subsec:temp_schedule_conve}

The energy and feedback-optimized methods have different performance characteristics.
The feedback method requires many sweeps per iteration in order to allow each replica 
to traverse between the extremal temperatures, but it generally converges after a few 
iterations. The energy method, on the other hand, can generate a good estimate of 
the energy values at each replica using much fewer sweeps per iteration compared to 
the feedback method, but its convergence is slow and depends on the initial temperature 
set.

 \begin{figure}[ht]
\begin{subfigure}[b]{0.49\linewidth}
\centering  
\includegraphics[width=1.05\linewidth]{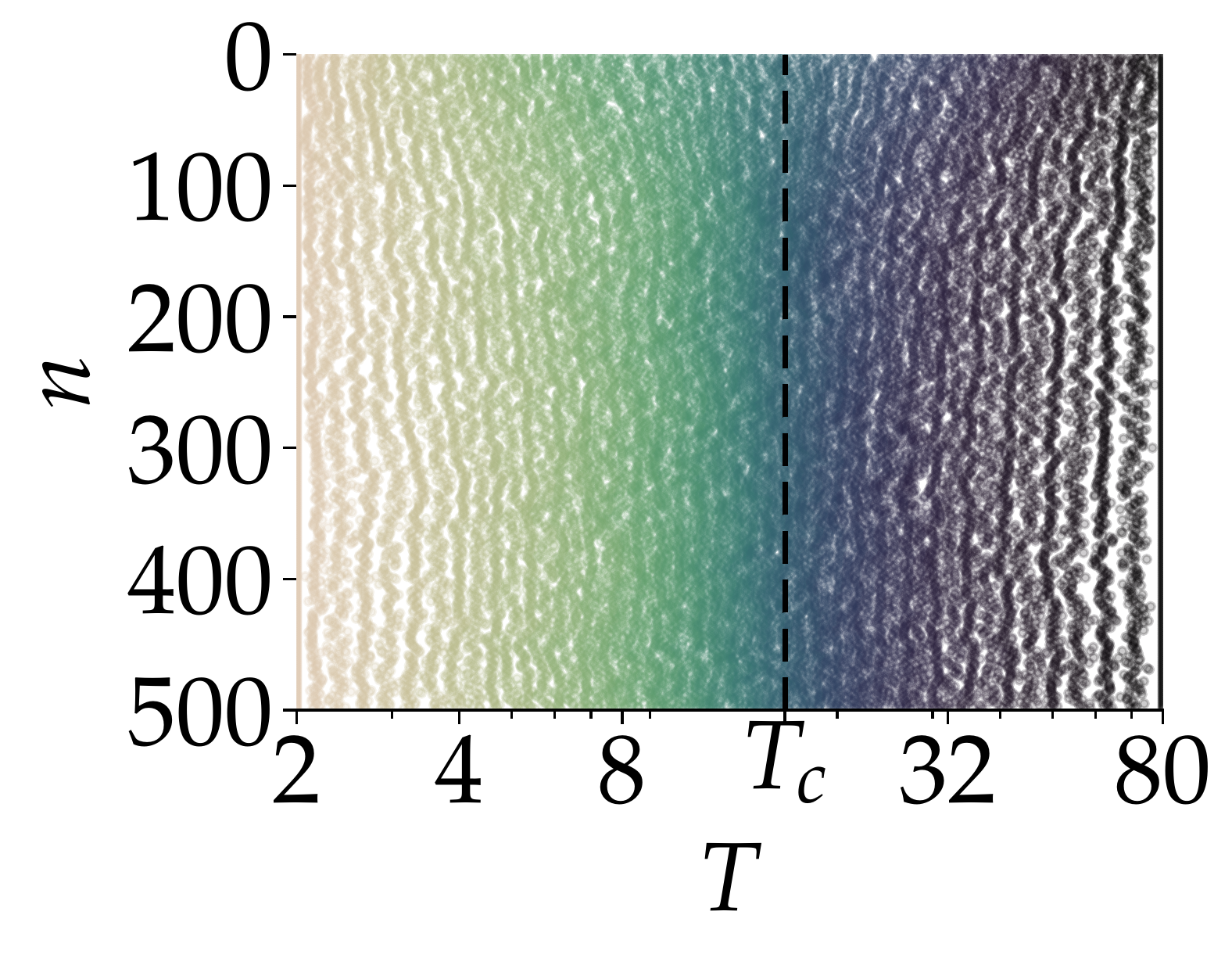}
\caption{\hspace{1.03cm}(a)}
\label{fig:energy_geometric}
\end{subfigure} 
\begin{subfigure}[b]{0.49\linewidth}
\centering
\includegraphics[width=1.05\linewidth]{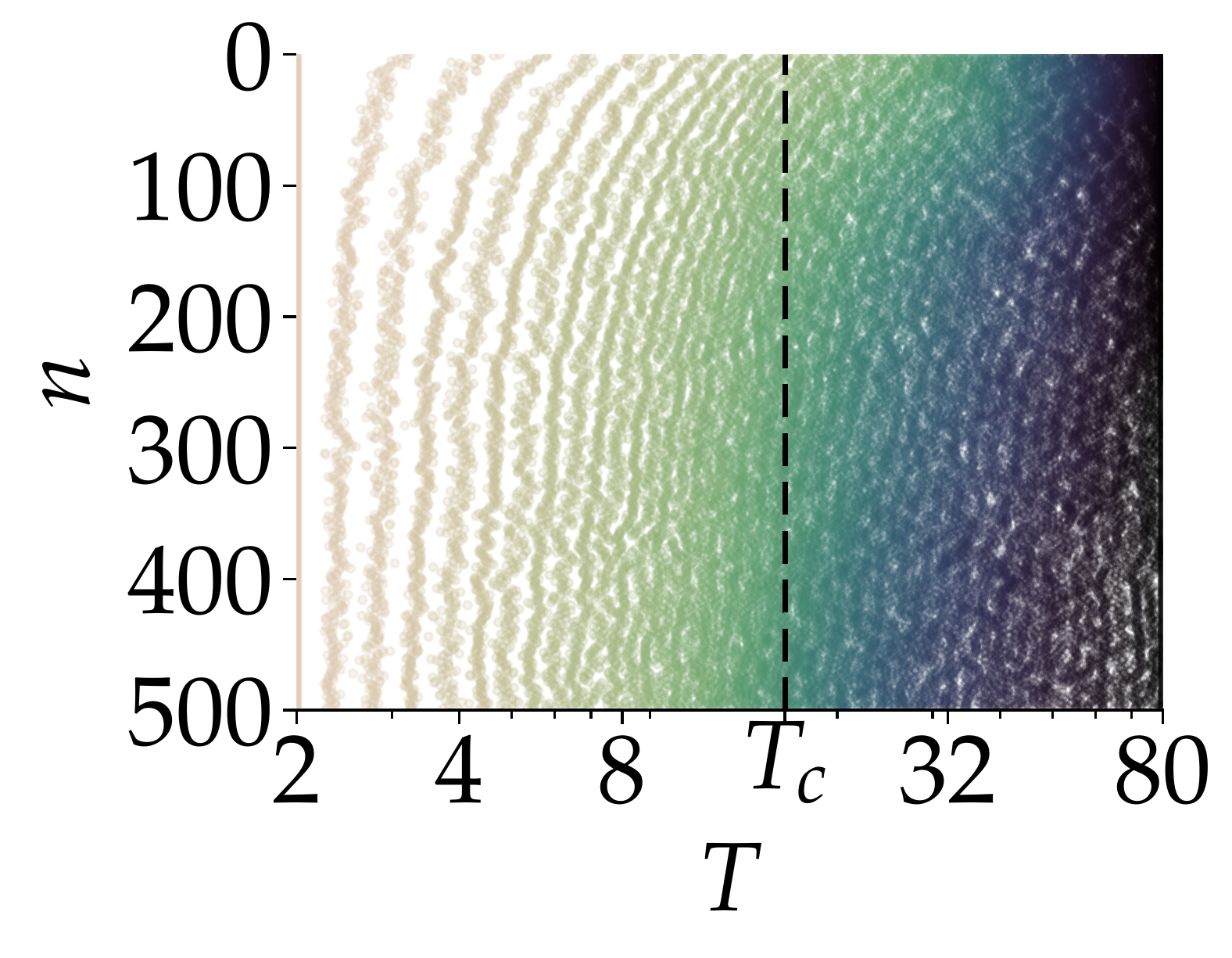}
\caption{\hspace{1.03cm}(b)}
\label{fig:energy_uniform}
\end{subfigure}
\begin{subfigure}[b]{0.49\linewidth}
\centering  
\includegraphics[width=1.05\linewidth]{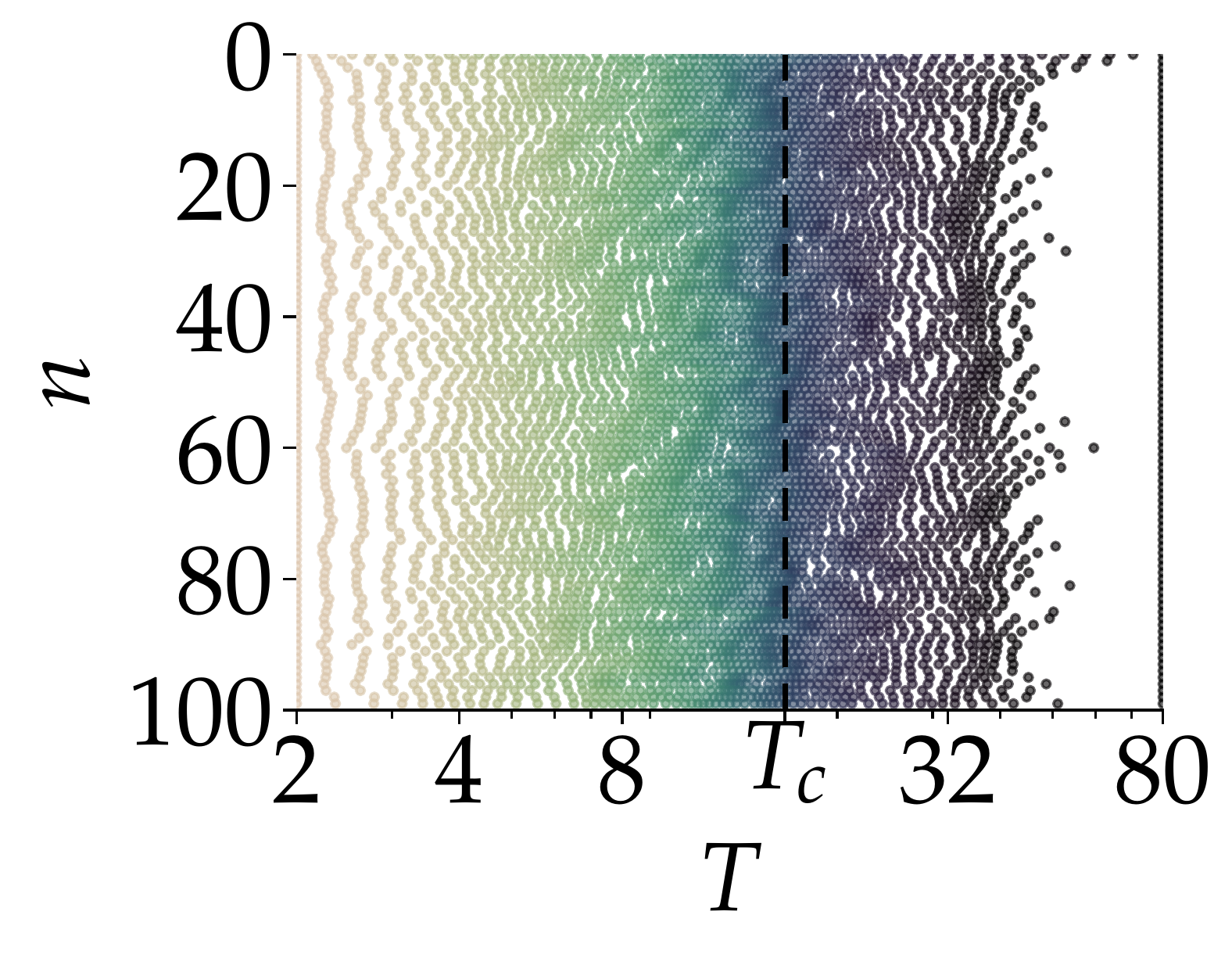}
\caption{\hspace{1.03cm}(c)}
\label{fig:feedback_geometric}
\end{subfigure}
\begin{subfigure}[b]{0.49\linewidth}
\centering
\includegraphics[width=1.05\linewidth]{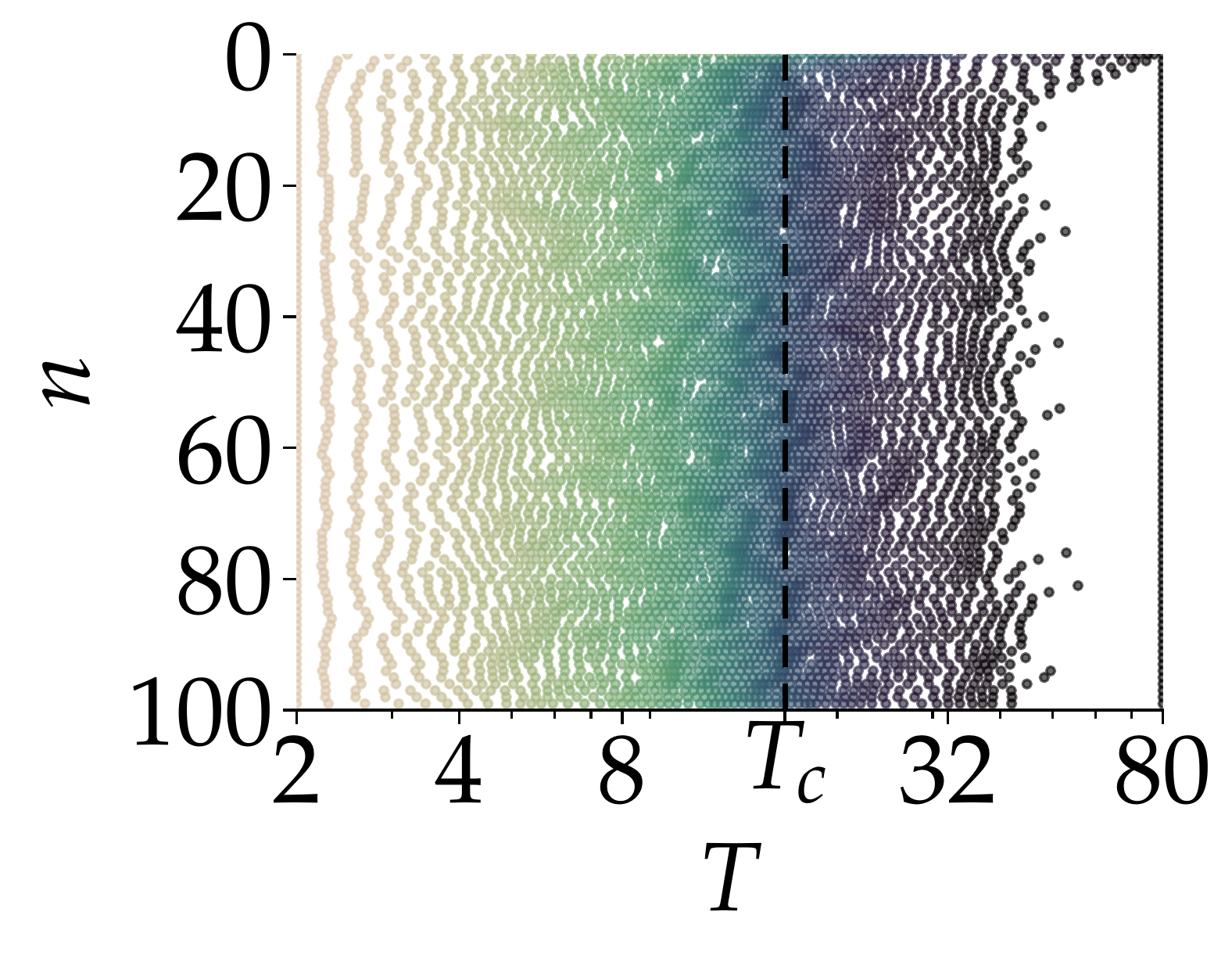}
\caption{\hspace{1.03cm}(d)}
\label{fig:feedback_uniform}
\end{subfigure}
\caption{Evolution of the temperature set through $500$ iterations of the 
energy method, starting from a geometric distribution in (a), and from 
a uniform distribution in (b); the evolution of the temperature set over 
$100$ iterations of the feedback-optimized method, starting from a geometric distribution 
in (c), and from a uniform distribution in (d). All experiments have been performed on an SK-bimodal 
problem instance of size $256$. The vertical dashed lines indicate the value of the 
critical temperature $T_c$ that roughly equals $\sqrt{N}$ for a fully connected 
spin-glass problem with bimodal disorder.}
\label{fig:convergence} 
\end{figure}

As illustrated in Fig.~\ref{fig:convergence}, it takes around $200$ iterations, each with $200$ 
sweeps, for the energy method to converge to a stable replica distribution when initialized 
from a geometric distribution, and even more iterations when starting from a uniform distribution. The 
feedback-optimized method, independent of the initial condition, converges after around five iterations, 
however, and each iteration performs $20\,000$ sweeps. It is worth noting that our experiments 
have shown that the feedback-optimized method is more prone to divergence. 
That is, after a number of iterations, the flow can become disconnected if one of the replicas fails to reach either
the highest or the lowest temperature. The energy method, in contrast, reliably converges given a sufficient   
number of iterations.

\subsection{Performance comparison}
\label{subsec:perf_comparison}
To quantify how the PT algorithm scales as the problem size increases using various 
replica distributions, we have measured the TTS. To observe the correct scaling of the PT algorithm, 
the minimum TTS for each problem class and size has been found by optimizing over the number 
of sweeps experimentally. More specifically, we have searched over different values of the number of sweeps, 
per problem class and size, to find the optimal number of sweeps that minimizes the TTS on 
a subset of problem instances. The TTS value, the acceptance probabilities of replica-exchange moves 
($\mathcal{P}_{\mbox{\scriptsize{RE}}}$), and the MC acceptance 
probabilities ($\mathcal{P}_{\mbox{\scriptsize{MC}}}$) have been then measured for each problem class
and size using all $100$ instances and the optimized number of sweeps.

\begin{figure*}[ht]
\begin{subfigure}[b]{0.5\linewidth}
\centering  
\includegraphics[width=1\linewidth]{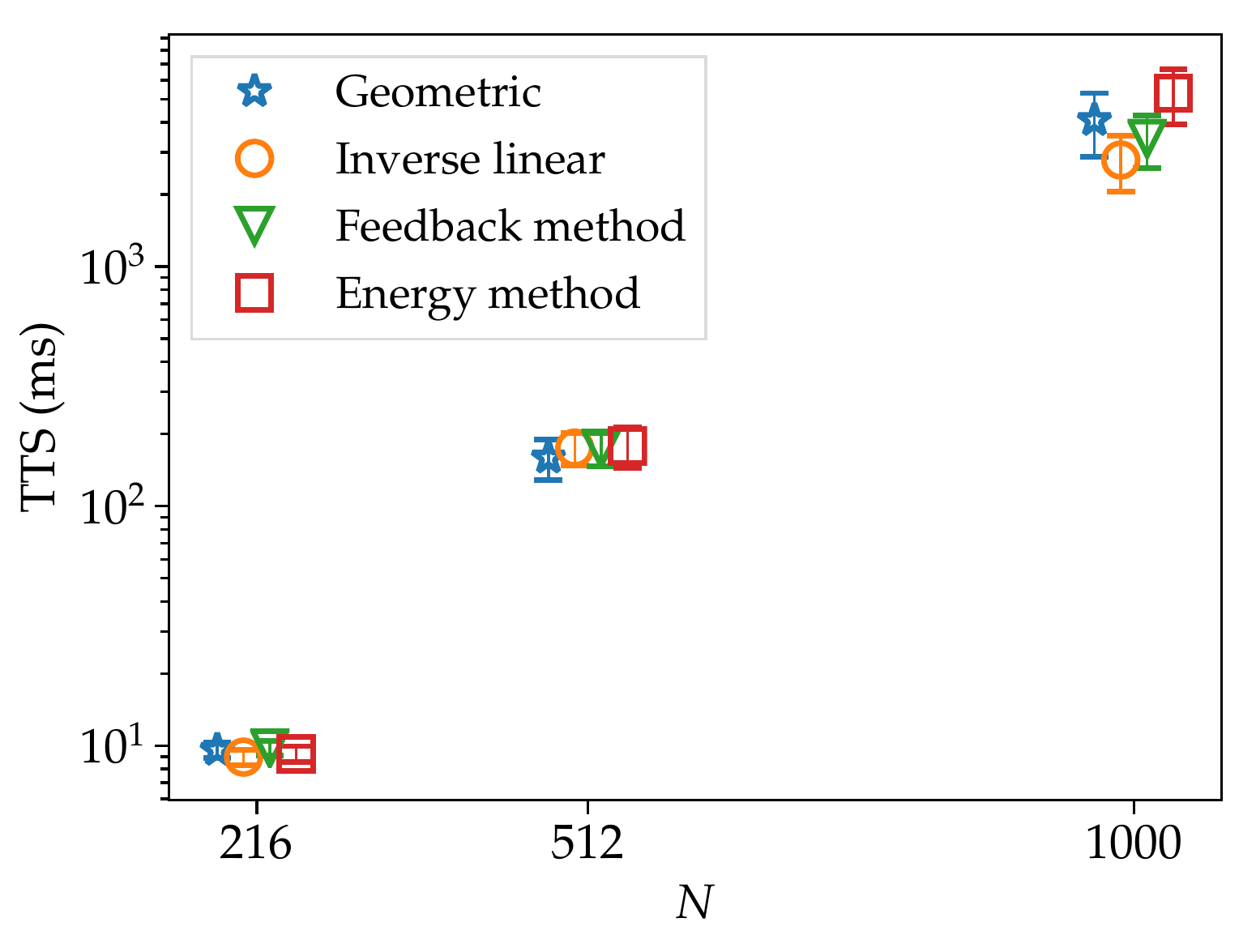} 
\caption{\hspace{0.9cm}(a)}
\label{fig:3D_Bimodal_tts_scaling}
\end{subfigure}
\begin{subfigure}[b]{0.5\linewidth}
\centering
\includegraphics[width=1\linewidth]{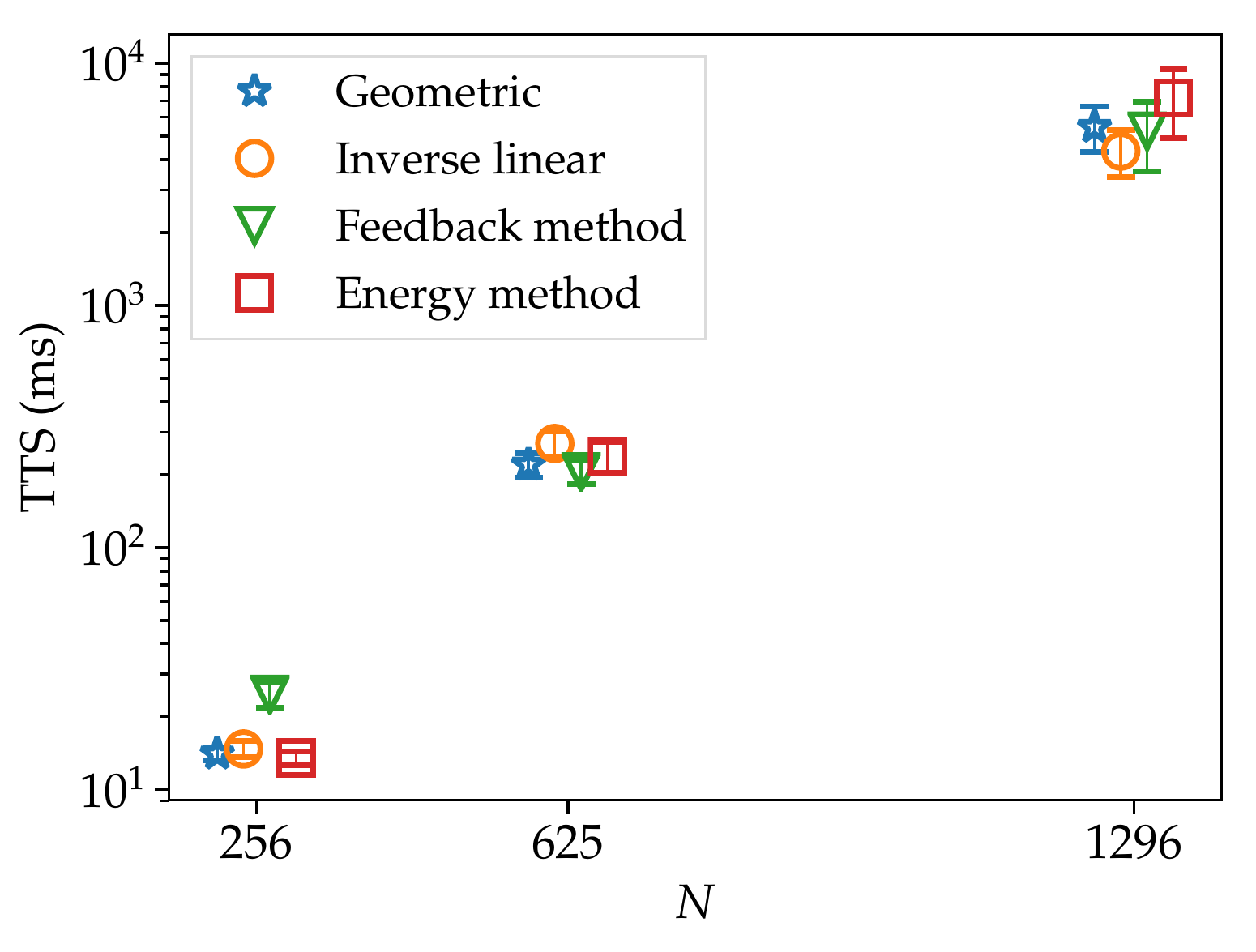}
\caption{\hspace{0.9cm}(b)}
\label{fig:4D_Bimodal_tts_scaling}
\end{subfigure}
\caption{The median TTS as a function of the number of variables $N$ for (a) 3D-bimodal 
and (b) 4D-bimodal spin-glass problems.}
\label{fig:tts_3D_4D} 
\end{figure*}

\vspace{5pt}
\paragraph*{Sparse Spin-Glass Problems} Figure~\ref{fig:tts_3D_4D} shows the mean and the standard 
deviation of the median TTS distribution of PT using four different
temperature setting methods on 3D and 4D problems. In all subsequent TTS plots, points and error
 bars represent the mean and the standard deviation of the distribution of the median TTS. 
 On both of these problem classes, the dynamic temperature setting methods 
do not outperform the static distributions for any of the three problem sizes. In terms of the
fraction of problems solved to the best-known energy, all methods have solved more than $95\%$ of the 
problems in each class for the first two sizes. For the largest size, the inverse-linear distribution 
has solved the most: 91\% and 98\% for 3D and 4D, respectively. Conversely, the energy method has 
solved the fewest problems to the best-known energy: 78\% and 89\% for 3D and 4D, respectively. 

\begin{figure}
\begin{subfigure}[b]{0.5\linewidth}
\centering  
\includegraphics[width=1\linewidth]{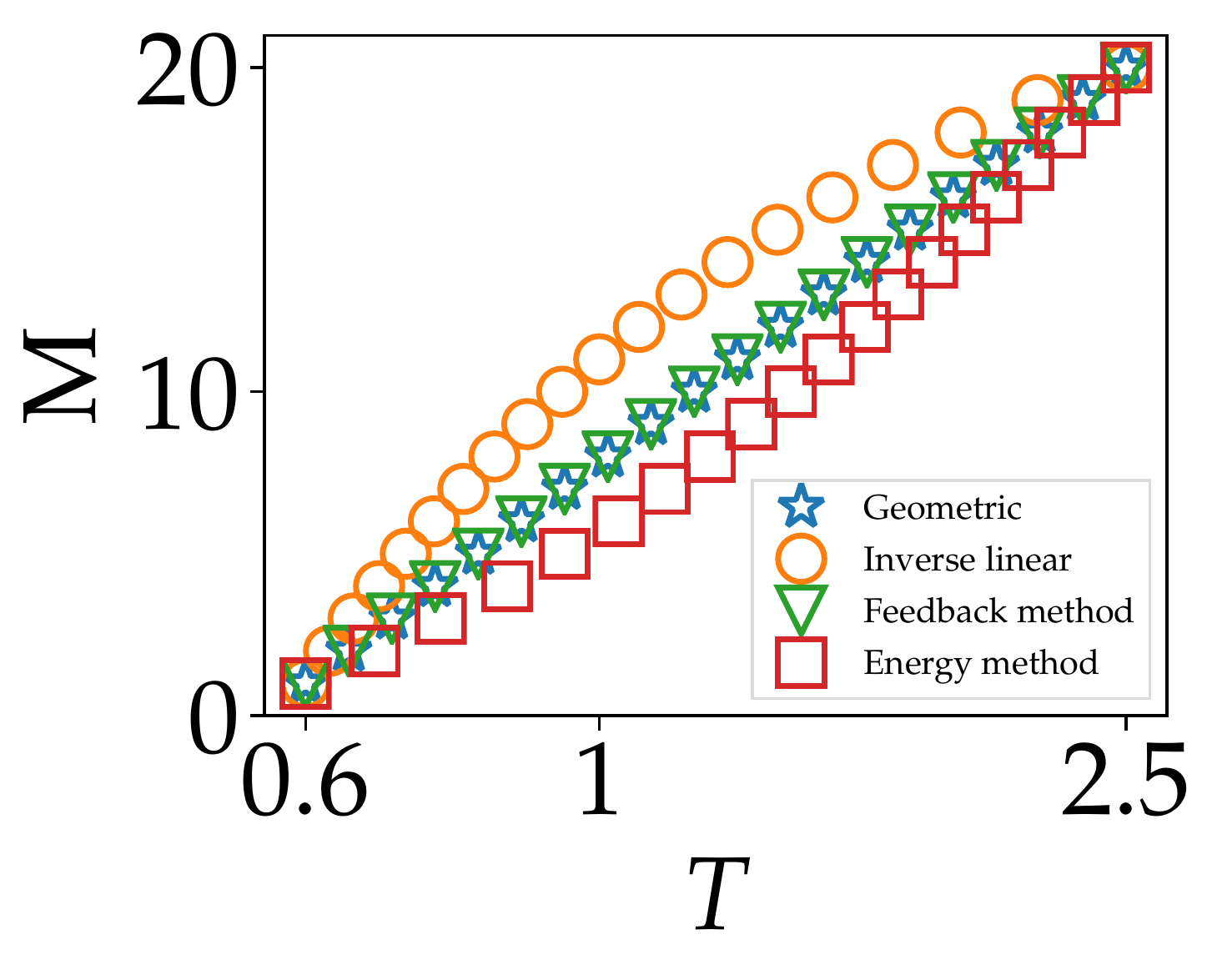} 
\caption{\hspace{0.8cm}(a)}
\label{fig:3D_Bimodal_temps}
\end{subfigure}
\begin{subfigure}[b]{0.5\linewidth}
\centering  
\includegraphics[width=1\linewidth]{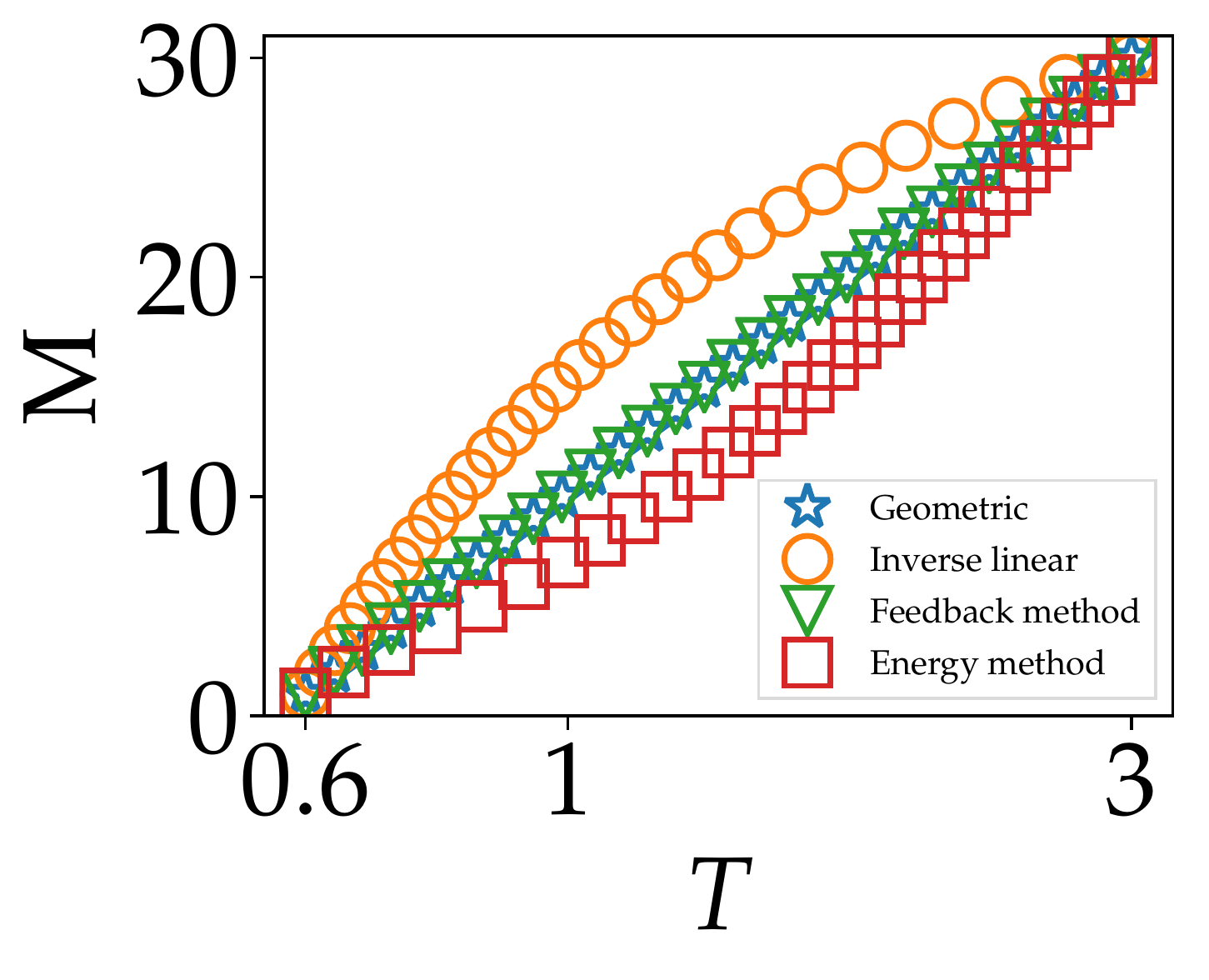} 
\caption{\hspace{0.8cm}(b)}
\label{fig:4D_Bimodal_temps}
\end{subfigure}
\caption{The cumulative number of temperatures below temperature $T$ (in log scale) 
obtained using four temperature setting methods for (a) a 3D-bimodal instance with $1000$ variables and 
for (b) a 4D-bimodal instance with $1296$ variables.}
\label{fig:detailed_temps_3D_4D} 
\end{figure}

\begin{figure}
\begin{subfigure}[b]{0.5\linewidth}
\centering
\includegraphics[width=1\linewidth]{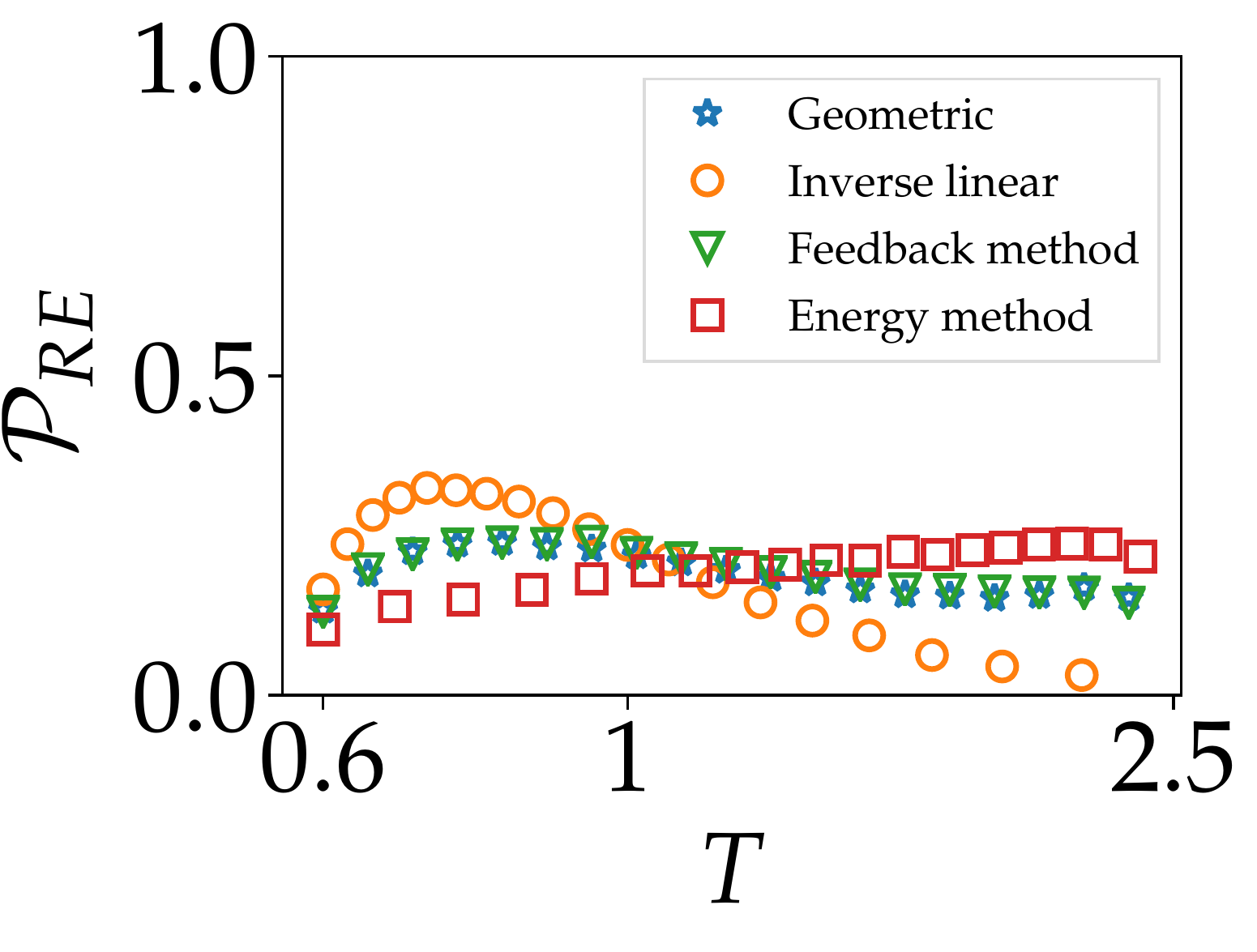}
\caption{\hspace{0.7cm}(a)}
\label{fig:3D_Bimodal_PT_probs}
\end{subfigure} 
\begin{subfigure}[b]{0.49\linewidth}
\centering
\includegraphics[width=1\linewidth]{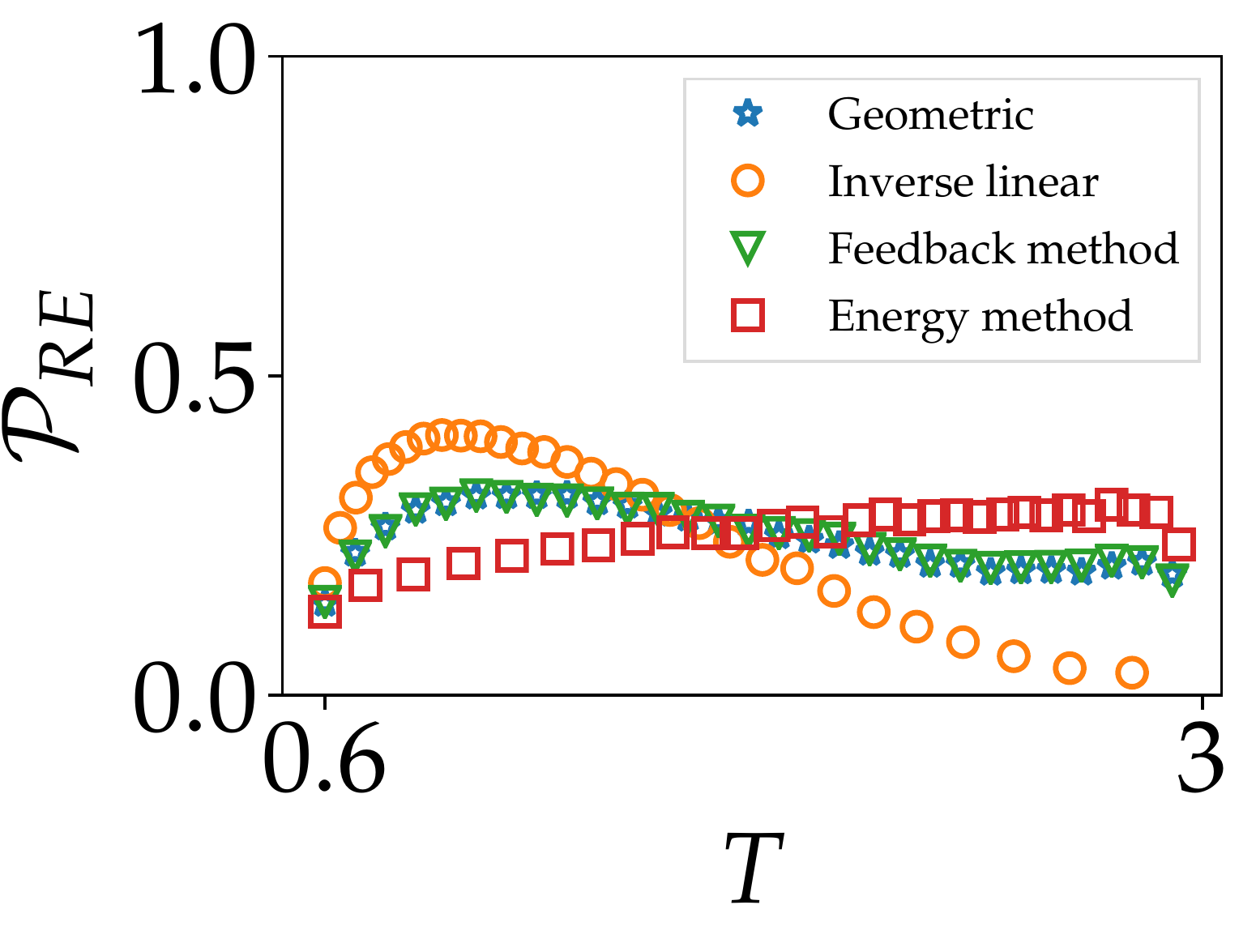}
\caption{\hspace{0.7cm}(b)}
\label{fig:4D_Bimodal_PT_probs}
\end{subfigure} 
\begin{subfigure}[b]{0.5\linewidth}
\centering
\includegraphics[width=1\linewidth]{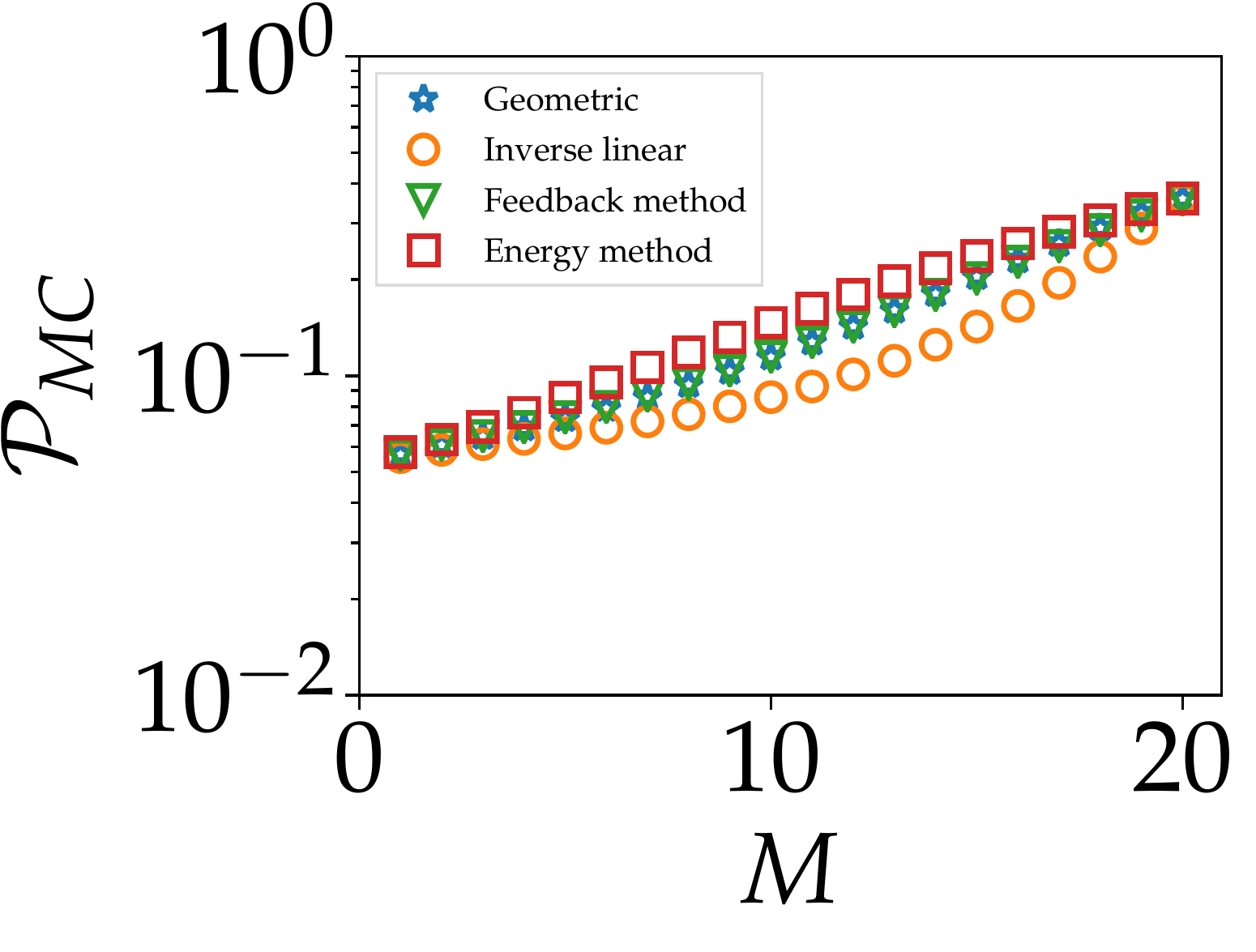}
\caption{\hspace{0.7cm}(c)}
\label{fig:3D_Bimodal_MC_probs}
\end{subfigure}
\begin{subfigure}[b]{0.5\linewidth}
\centering
\includegraphics[width=1\linewidth]{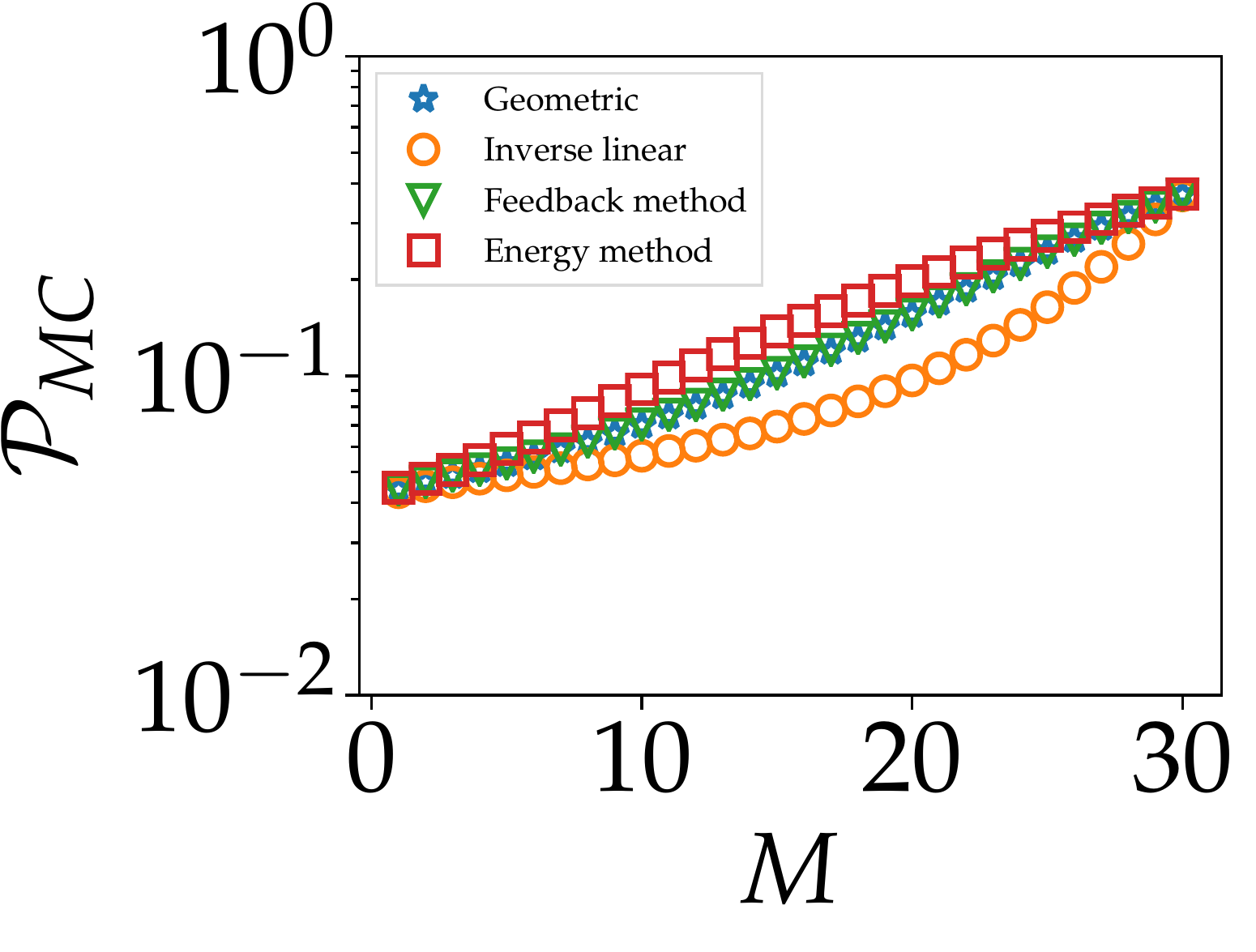}
\caption{\hspace{0.7cm}(d)}
\label{fig:4D_Bimodal_MC_probs}
\end{subfigure}
\caption{The acceptance probabilities of replica-exchange moves 
($\mathcal{P}_{\mbox{\scriptsize{RE}}}$), and the MC acceptance 
probabilities ($\mathcal{P}_{\mbox{\scriptsize{MC}}}$) 
obtained using four methods for a 3D-bimodal instance with $1000$ 
variables are shown in (a) and (c), respectively. The same measurements are illustrated, 
in respective terms, for a 4D-bimodal instance with $1296$ variables in (b) and (d). The 
standard deviations for the MC probabilities are very small, and so have not been included. 
Those for replica-exchange probabilities were also excluded, but in this case for better readability.}
\label{fig:detailed_3D_4D} 
\end{figure}

Figures~\ref{fig:detailed_temps_3D_4D} and~\ref{fig:detailed_3D_4D} illustrate 
the temperature distribution and the acceptance probabilities including the 
replica-exchange probability ($\mathcal{P}_{\mbox{\scriptsize{RE}}}$) at each replica, 
and the MC acceptance probability ($\mathcal{P}_{\mbox{\scriptsize{MC}}}$) at each replica 
for an instance of the largest problem size in each problem category of 3D-bimodal and 4D-bimodal. 

\begin{figure*}[ht!]
\begin{subfigure}[b]{0.5\linewidth}
\centering  
\includegraphics[width=1\linewidth]{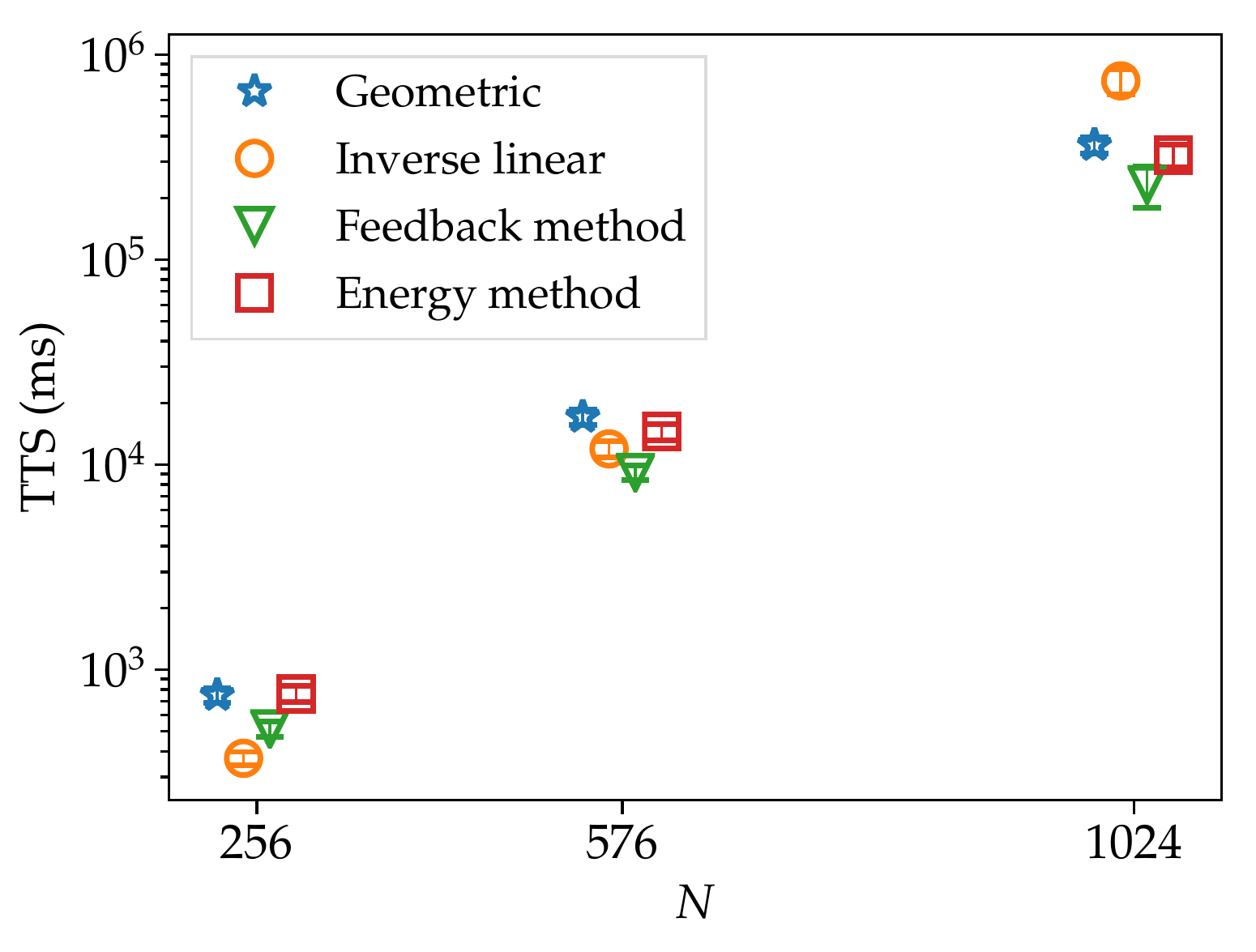} 
\caption{\hspace{0.9cm}(a)}
\label{fig:3D_Bimodal_tts_scaling}
\end{subfigure}
\begin{subfigure}[b]{0.5\linewidth}
\centering
\includegraphics[width=1\linewidth]{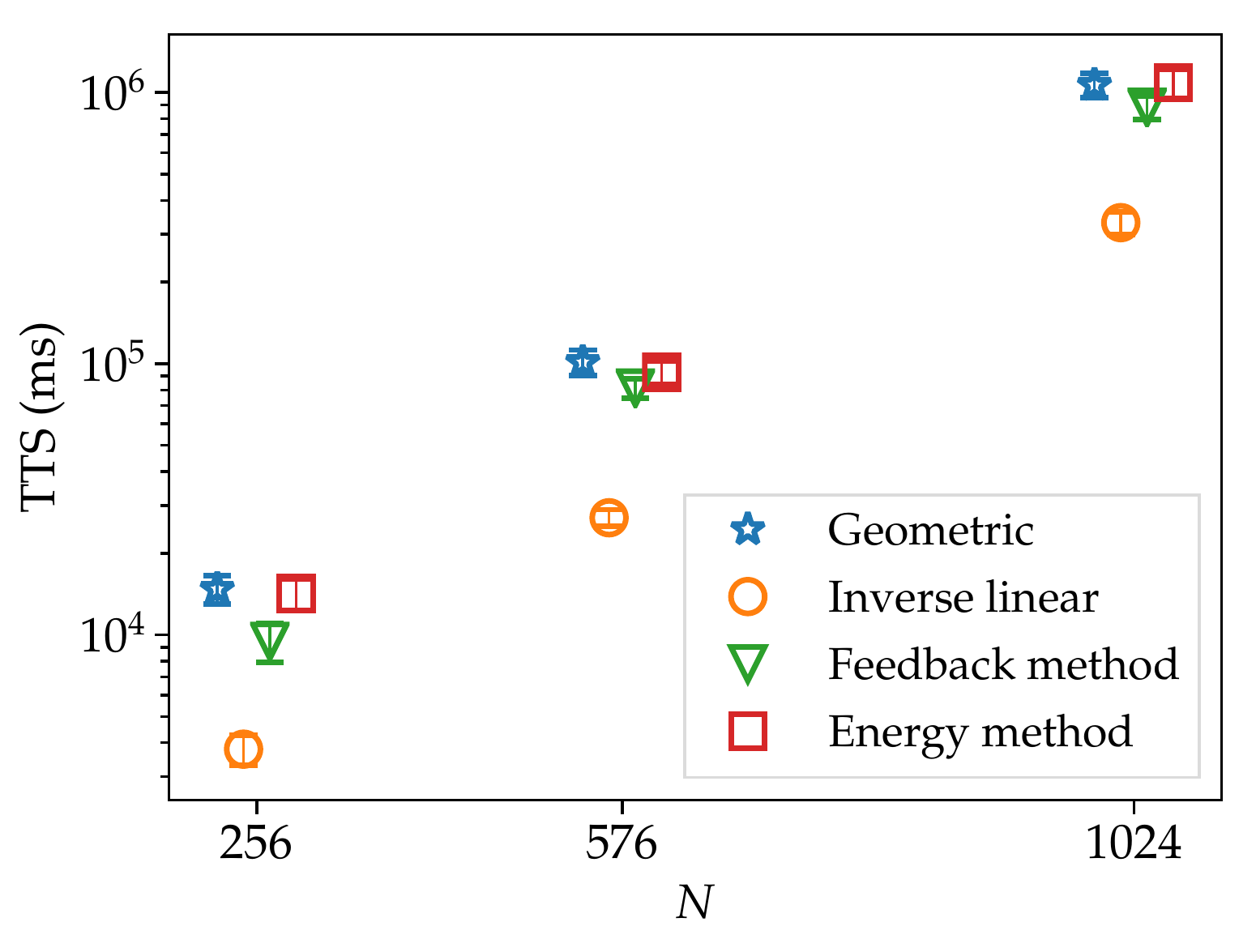}
\caption{\hspace{0.9cm}(b)}
\label{fig:4D_Bimodal_tts_scaling}
\end{subfigure}
\caption{The median TTS of fully connected SK spin-glass problems with (a) bimodal and (b) Gaussian disorder versus 
the number of variables $N$.}
\label{fig:tts_SK} 
\end{figure*}

Figure~\ref{fig:detailed_temps_3D_4D} illustrates that the feedback-optimized method does not 
improve on its initial temperature schedule, that is, the geometric distribution. In other words, the geometric 
distribution on sparse spin-glass problems turns out to have a closer flow to the optimal flow. 
Consequently, as shown in Fig.~\ref{fig:detailed_3D_4D}, both feedback-optimized and 
geometric temperature schedules demonstrate similar $\mathcal{P}_{\mbox{\scriptsize{RE}}}$ 
and $\mathcal{P}_{\mbox{\scriptsize{MC}}}$ distributions. The energy method, in contrast, 
changes its initial distribution: It has arranged fewer temperature values around the minimum 
temperature value, which results in a more uniform $\mathcal{P}_{\mbox{\scriptsize{RE}}}$ 
distribution, and has higher $\mathcal{P}_{\mbox{\scriptsize{MC}}}$ values. 
Having fewer temperatures close to the lowest temperature could indicate that the 
diffusion process that distributes replicas could require a much longer time to 
converge to the schedule with uniform swap rate. This may explain the slightly 
worse performance of the energy method for the largest problem size compared 
to the other three methods. The small performance improvement of the inverse-linear temperature method is likely due to the fact 
that its temperature distribution has the highest density around the minimum temperature value, 
its $\mathcal{P}_{\mbox{\scriptsize{RE}}}$ distribution has a large peak at low temperature values, 
and its $\mathcal{P}_{\mbox{\scriptsize{MC}}}$ values are the lowest.

\vspace{5pt}
\paragraph*{Fully connected spin-glass problems} 
Figure~\ref{fig:tts_SK} illustrates the mean and the standard deviation 
of the median TTS of PT using four temperature setting methods on 
the SK-bimodal and SK-Gaussian problem classes. Similarly to the 
sparse spin-glass problems, none of the dynamic temperature setting methods 
shows a clear advantage over the static replica distribution methods. The inverse-linear 
temperature set, however, has noticeably superior performance compared to the 
other methods for the SK-Gaussian problem class. 

\begin{figure}
\begin{subfigure}[b]{0.5\linewidth}
\centering  
\begin{overpic}[width=1\linewidth]{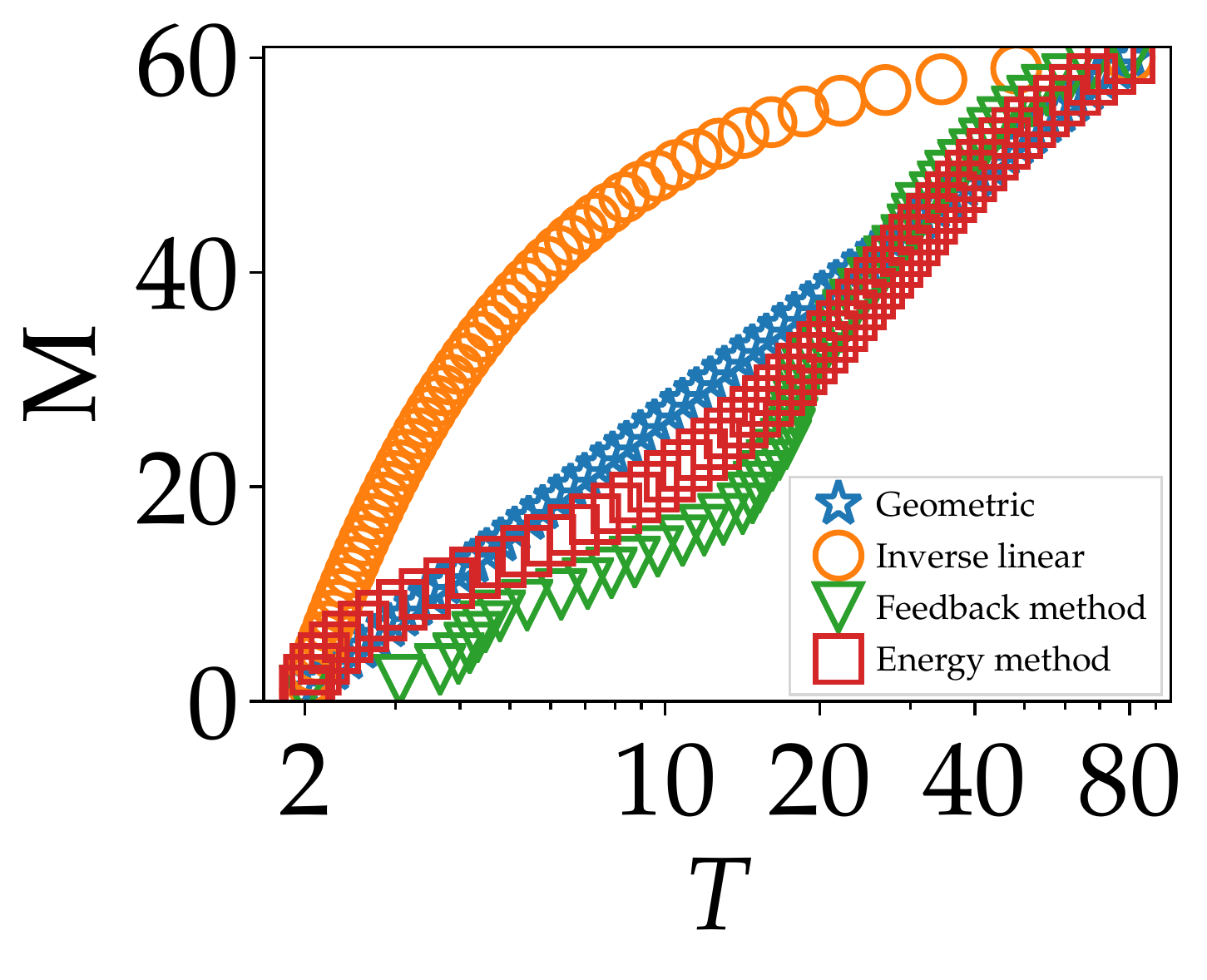} 
\end{overpic}
\caption{(a)}
\label{fig:SK_Bimodal_temps}
\end{subfigure}
\begin{subfigure}[b]{0.5\linewidth}
\centering  
\begin{overpic}[width=1\linewidth]{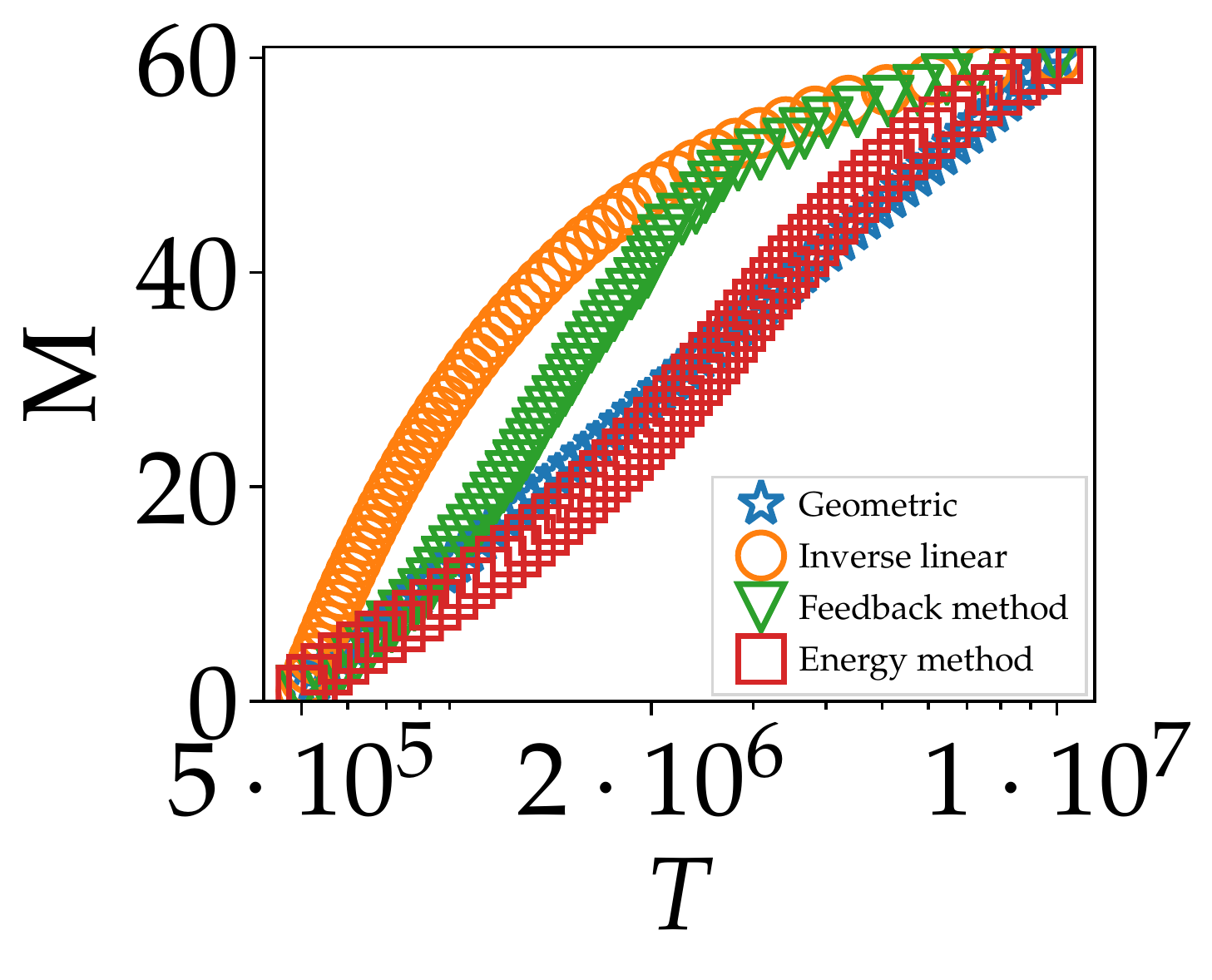} 
\end{overpic}
\caption{(b)}
\label{fig:SK_Gaussian_temps}
\end{subfigure}
\caption{The cumulative distribution of number of temperatures obtained for a fully connected 
SK spin-glass problem instance with $1024$ variables with (a) bimodal and 
(b) Gaussian couplings using different temperature setting methods. The temperature 
values shown on the $x$-axis are in log scale.}
\label{fig:detailed_temps_SK} 
\end{figure}

\begin{figure}
\begin{subfigure}[b]{0.5\linewidth}
\centering
\includegraphics[width=1\linewidth]{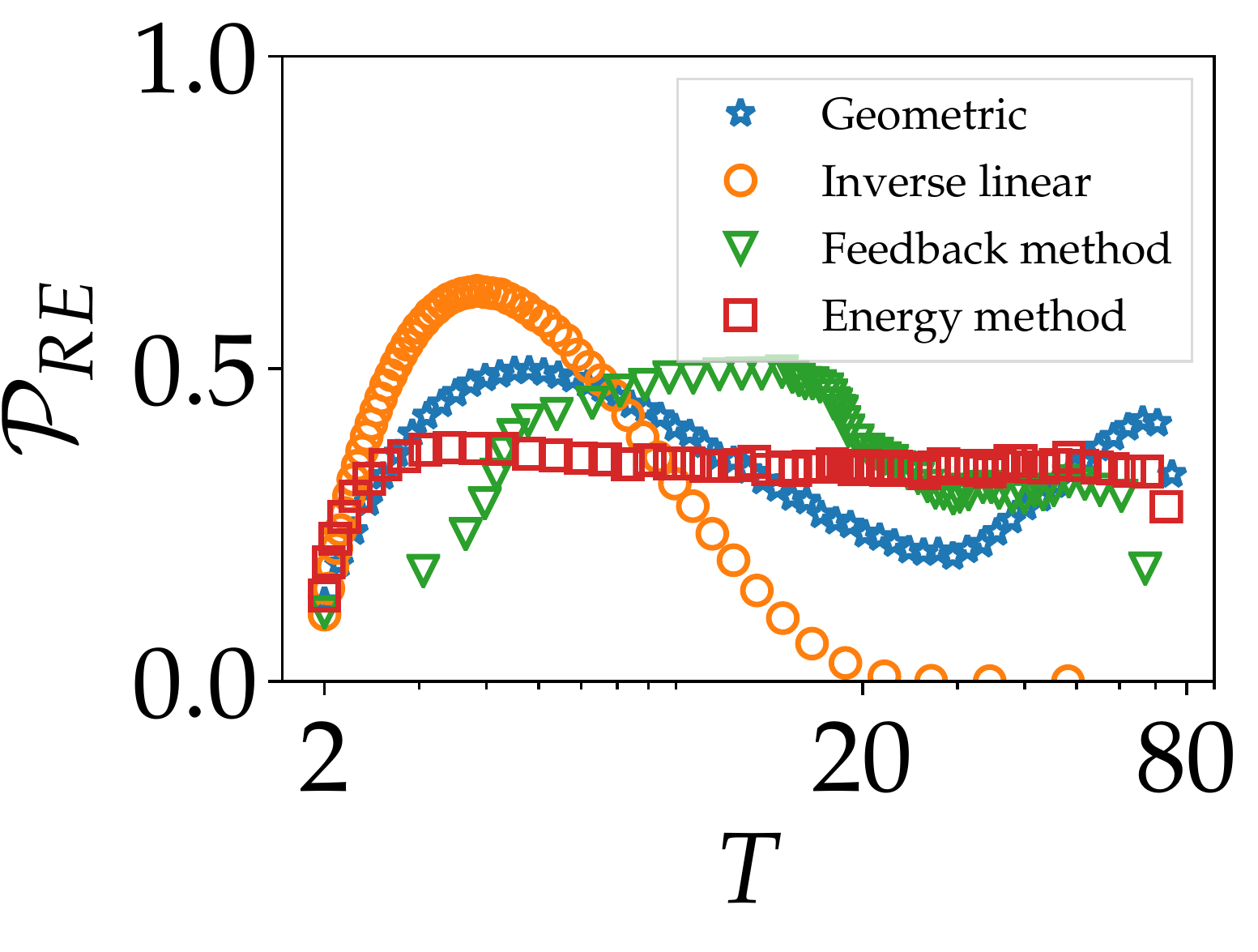}
\caption{\hspace{0.7cm}(a)}
\label{fig:SK_Bimodal_PT_probs}
\end{subfigure} 
\begin{subfigure}[b]{0.49\linewidth}
\centering
\includegraphics[width=1\linewidth]{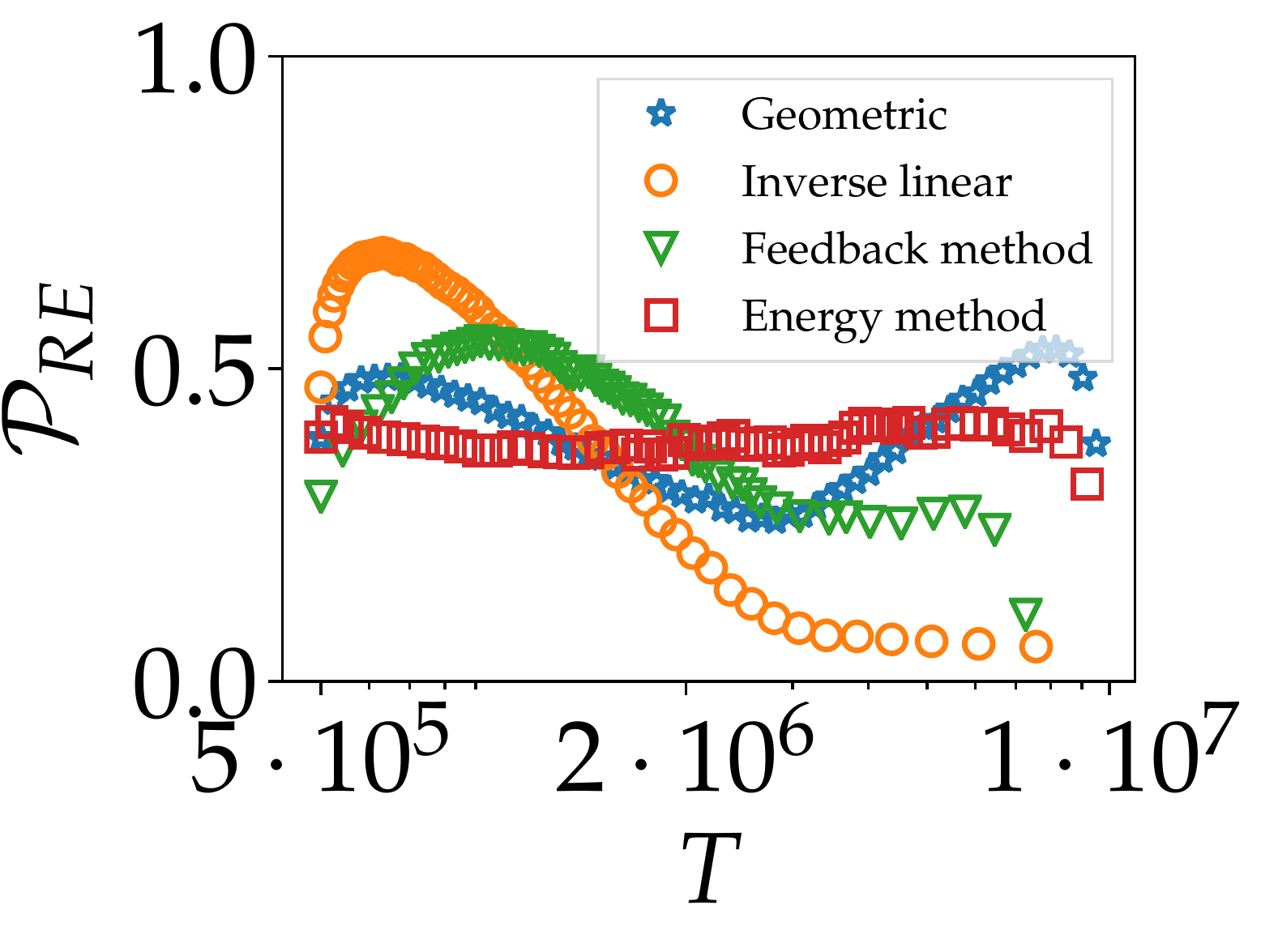}
\caption{\hspace{0.7cm}(b)}
\label{fig:SK_Gaussian_PT_probs}
\end{subfigure} 
\begin{subfigure}[b]{0.5\linewidth}
\centering
\includegraphics[width=1\linewidth]{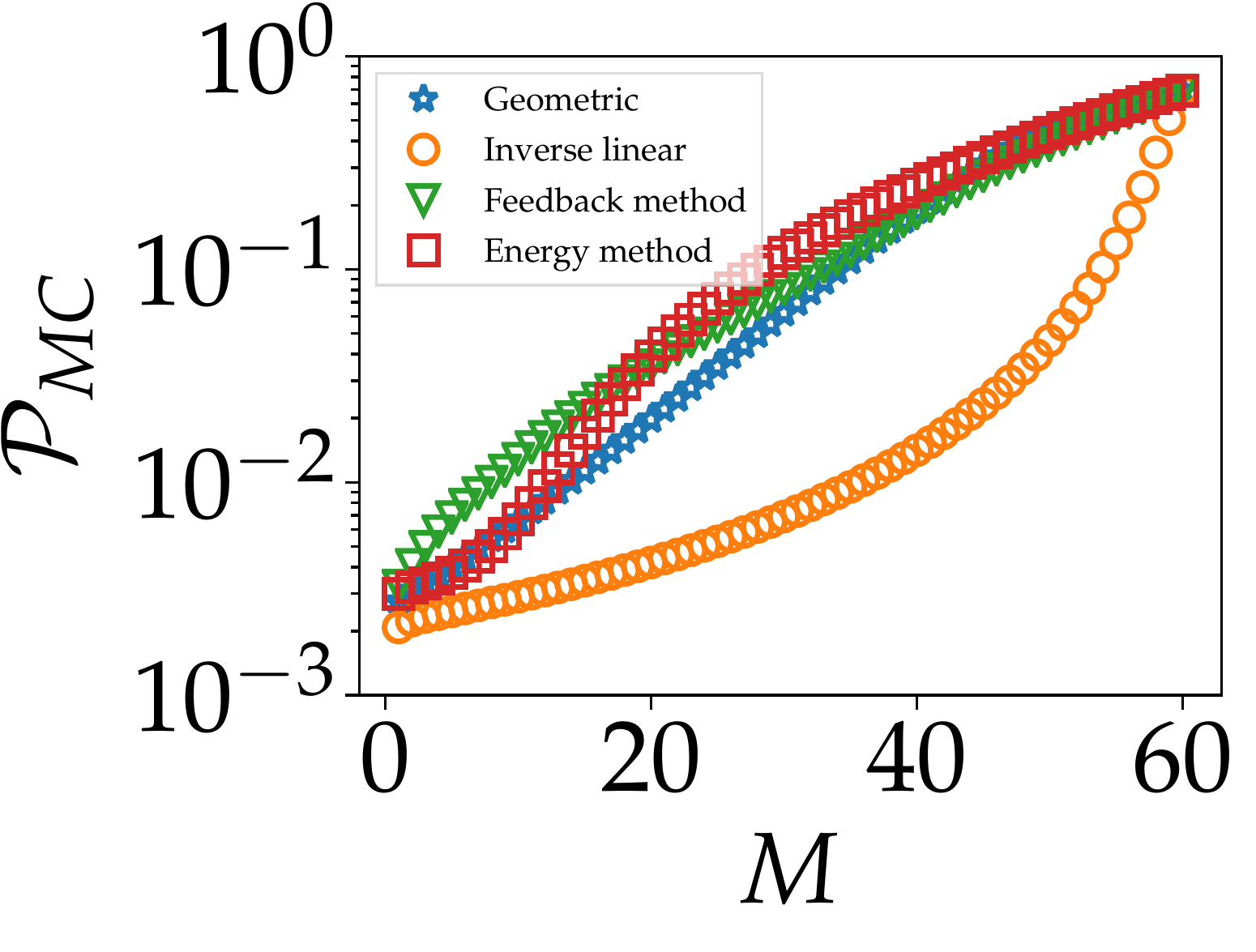}
\caption{\hspace{0.7cm}(c)}
\label{fig:SK_Bimodal_MC_probs}
\end{subfigure}
\begin{subfigure}[b]{0.5\linewidth}
\centering
\includegraphics[width=1\linewidth]{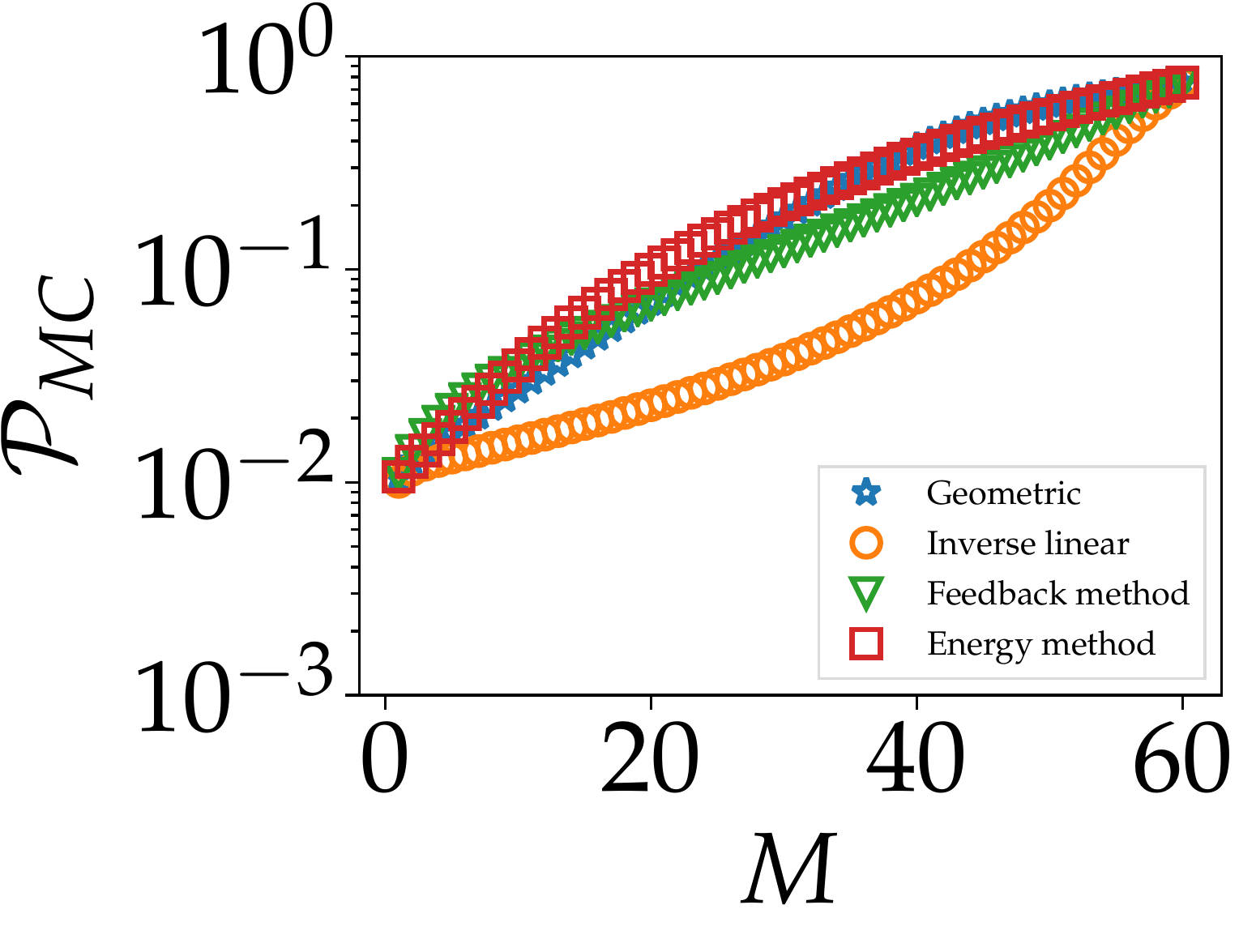}
\caption{\hspace{0.7cm}(d)}
\label{fig:SK_Gaussian_MC_probs}
\end{subfigure}
\caption{Replica-exchange ($\mathcal{P}_{\mbox{\scriptsize{RE}}}$) and  
MC ($\mathcal{P}_{\mbox{\scriptsize{MC}}}$) acceptance probability distributions of an 
instance of the {SK-bimodal} problem class with $1024$ variables are shown in (a) and (c), respectively, 
for different temperature setting methods. The same measurements are illustrated, 
in respective terms, for an SK-Gaussian problem instance with $1024$ variables in (b) and (d). }
\label{fig:detailed_SK} 
\end{figure}

As shown in Fig.~\ref{fig:detailed_temps_SK}, for both the SK-bimodal and SK-Gaussian problem classes, the inverse-linear temperature 
schedule has arranged more temperature values close to the minimum temperature, resulting
in similar $\mathcal{P}_{\mbox{\scriptsize{RE}}}$ and $\mathcal{P}_{\mbox{\scriptsize{MC}}}$ 
distributions. However, for the former class, the inverse-linear schedule yields poor performance, whereas 
for the latter, it results in a performance advantage. This behavioural difference is likely due to the fact 
that the replica-exchange probabilities for SK-bimodal problem instances are very close to zero 
for the replicas associated with temperature values between $20$ and $80$ (Fig.~\ref{fig:detailed_SK}). 
Having near-zero exchange probabilities at high temperature values does not allow for a continuous reseeding of 
the states in the PT algorithm. In contrast, the replica-exchange probability values for SK-Gaussian 
problem instances are much higher over the same temperature range.

Our results for sparse and fully connected spin-glass problems that have continuous phase transitions 
in the order parameter do not show an advantage for the feedback-optimized method. The base inverse-linear 
method that places more temperature values around the minimum temperature results in better performance 
at a much lower computational cost. 

\begin{figure}[ht!]
\centering  
\begin{overpic}[width=1\linewidth]{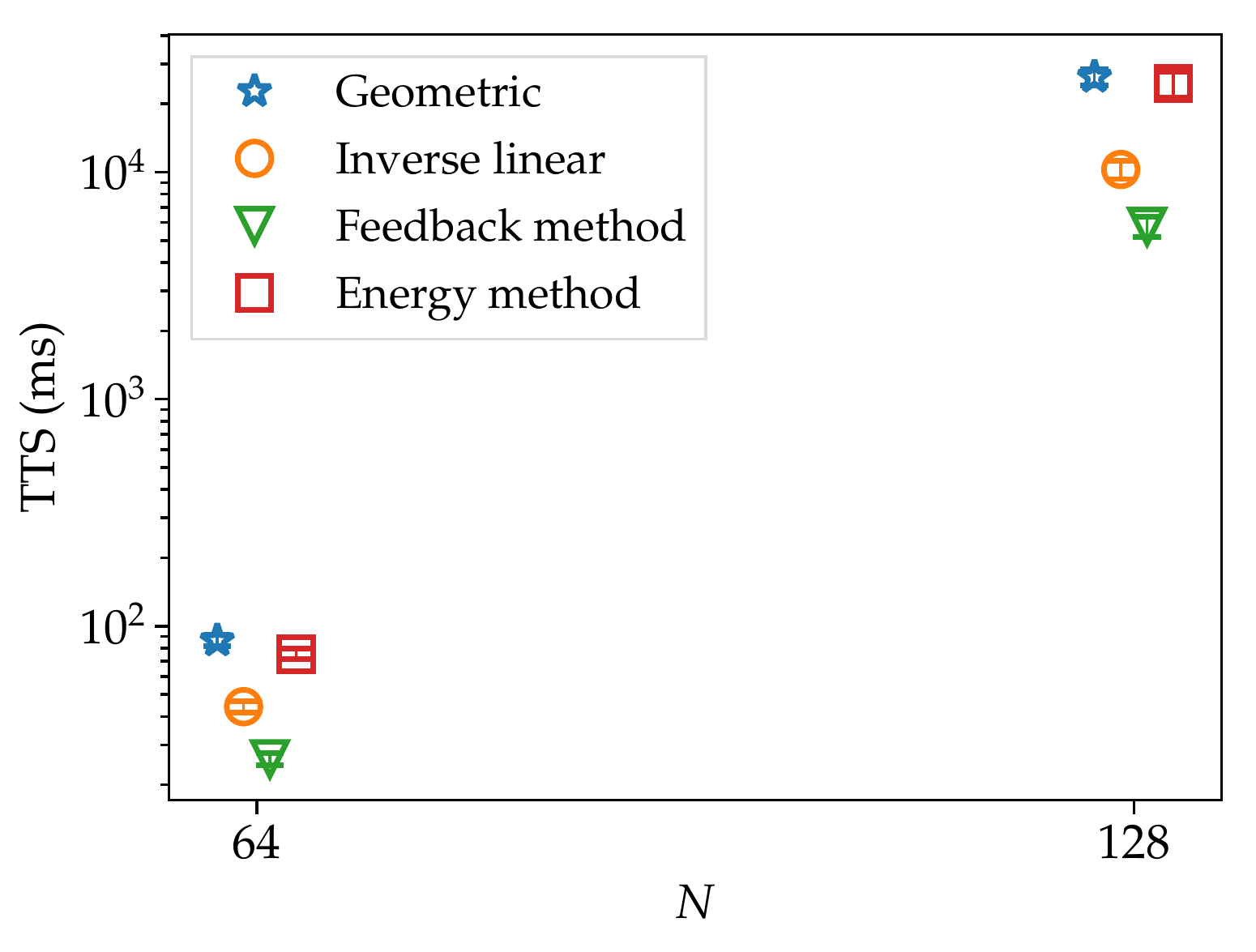} 
\end{overpic}
\caption{The median TTS of fully connected Wishart problems versus the number of variables $N$.}
\label{fig:tts_Wishart} 
\end{figure}

\paragraph*{Fully connected Wishart problems}
The mean and the standard deviation of the median TTS for the PT algorithm for two relatively small-sized Wishart problems are 
shown in Fig.~\ref{fig:tts_Wishart}. The dynamic feedback-optimized temperature setting method
exhibits a TTS that is approximately two times faster than the inverse-linear method and
roughly five times faster than the energy and the geometric methods. Due to the inherent computational 
difficulty built into Wishart problems, we have been unable to solve a sufficient number of larger Wishart 
problem instances to optimality in order to measure the median TTS for bigger problem sizes. 

\begin{figure}
\centering   
\begin{overpic}[width=1\linewidth]{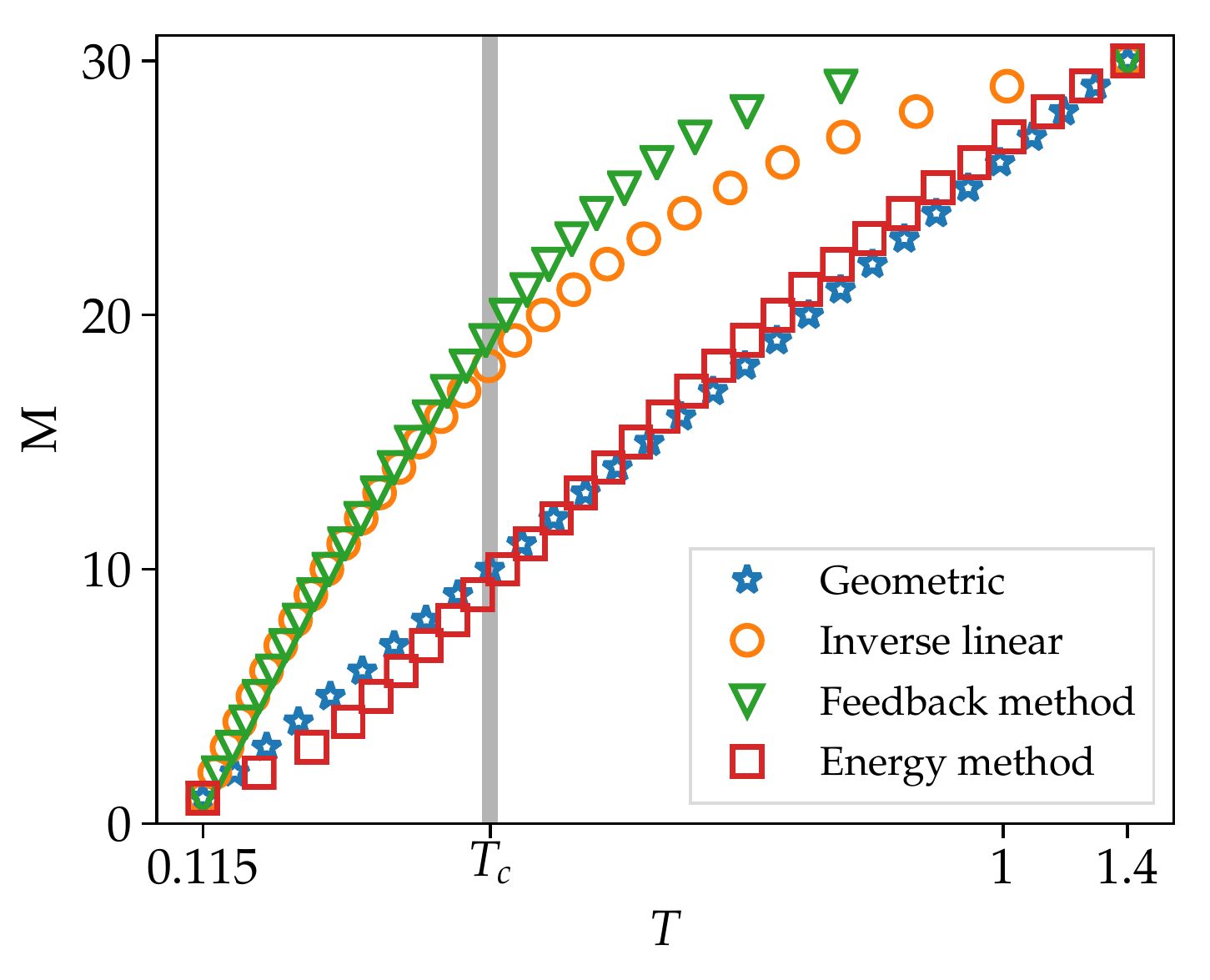} 
\end{overpic}
\caption{The cumulative distribution of number of temperatures obtained for a fully connected Wishart problem 
instance of size $128$ using four temperature setting methods. The temperature 
values shown on the $x$ axis are in log scale.}
\label{fig:detailed_temps_Wishart} 
\end{figure}

\begin{figure}
\begin{subfigure}[b]{0.49\linewidth}
\centering
\includegraphics[width=1\linewidth]{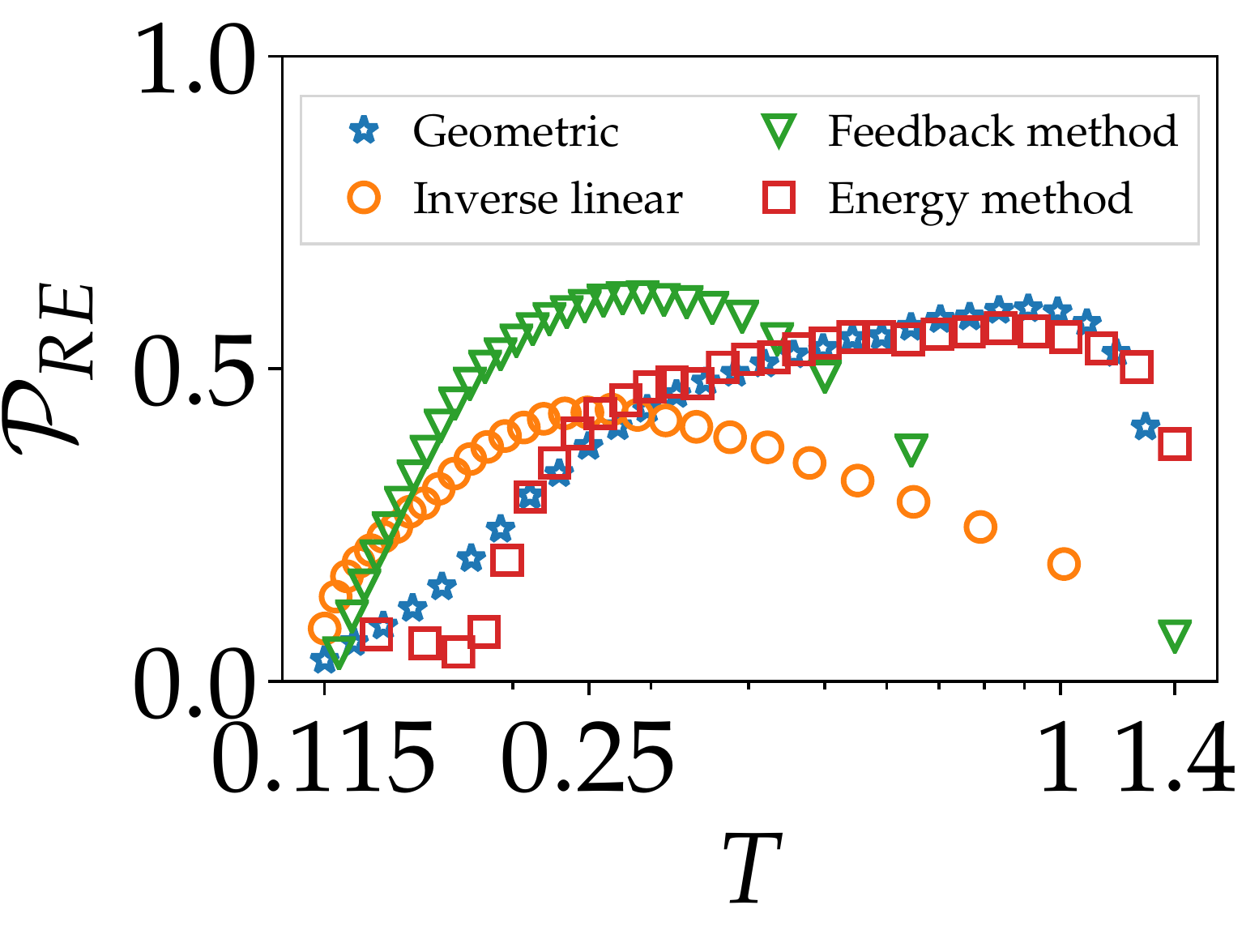}
\caption{\hspace{0.7cm}(a)}
\label{fig:SK_Wishart_PT_probs}
\end{subfigure} 
\begin{subfigure}[b]{0.49\linewidth}
\centering
\includegraphics[width=1\linewidth]{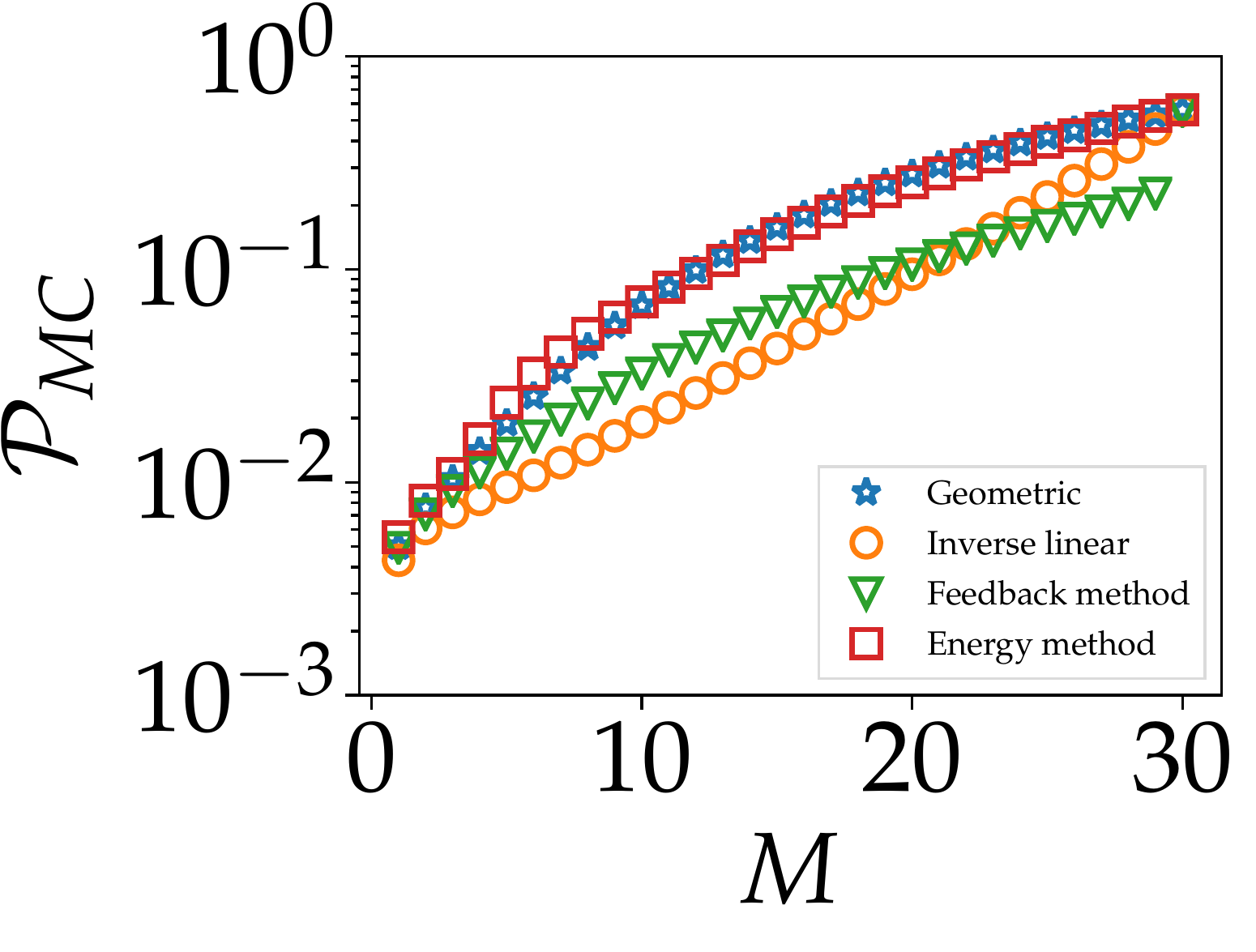}
\caption{\hspace{0.7cm}(b)}
\label{fig:SK_Wishart_MC_probs}
\end{subfigure}
\caption{Acceptance probabilities for replica-exchange moves 
($\mathcal{P}_{\mbox{\scriptsize{RE}}}$) and MC acceptance probabilities 
($\mathcal{P}_{\mbox{\scriptsize{MC}}}$) obtained using four methods 
for a Wishart instance of size $128$ are shown in (a) and (b), respectively.}
\label{fig:detailed_Wishart} 
\end{figure}

A closer look into the results shown in Fig.~\ref{fig:detailed_temps_Wishart} and 
Fig.~\ref{fig:detailed_Wishart} demonstrates that, for Wishart problems, the energy method essentially 
converges to its initial geometric schedule, whereas the feedback method and the inverse-linear methods 
place twice as many replicas at or below the critical temperature. As mentioned in Sec.~\ref{subsec:energy}, the energy 
method is designed to have equal replica-exchange probabilities for all adjacent temperature values. 
Despite the roughly uniform $\mathcal{P}_{\mbox{\scriptsize{RE}}}$ distribution 
of the energy method for the previous two problem classes, the energy method results in a 
nonuniform distribution for Wishart problem instances. 

In contrast to spin-glass problems, the Wishart problems have a first-order 
phase transition.  As shown in Fig.~\ref{fig:detailed_Wishart}, the feedback-optimized 
method distributes the temperature values such that the replica-exchange probabilities 
are the highest around the critical temperature ($T_c \simeq 0.25$), compared to the other 
temperature setting methods. This maximizes the temperature mixing around the critical 
temperature and consequently yields a smaller value for the TTS.

\begin{figure}[h]
\begin{subfigure}[b]{0.49\linewidth}
\centering
\includegraphics[width=1\linewidth]{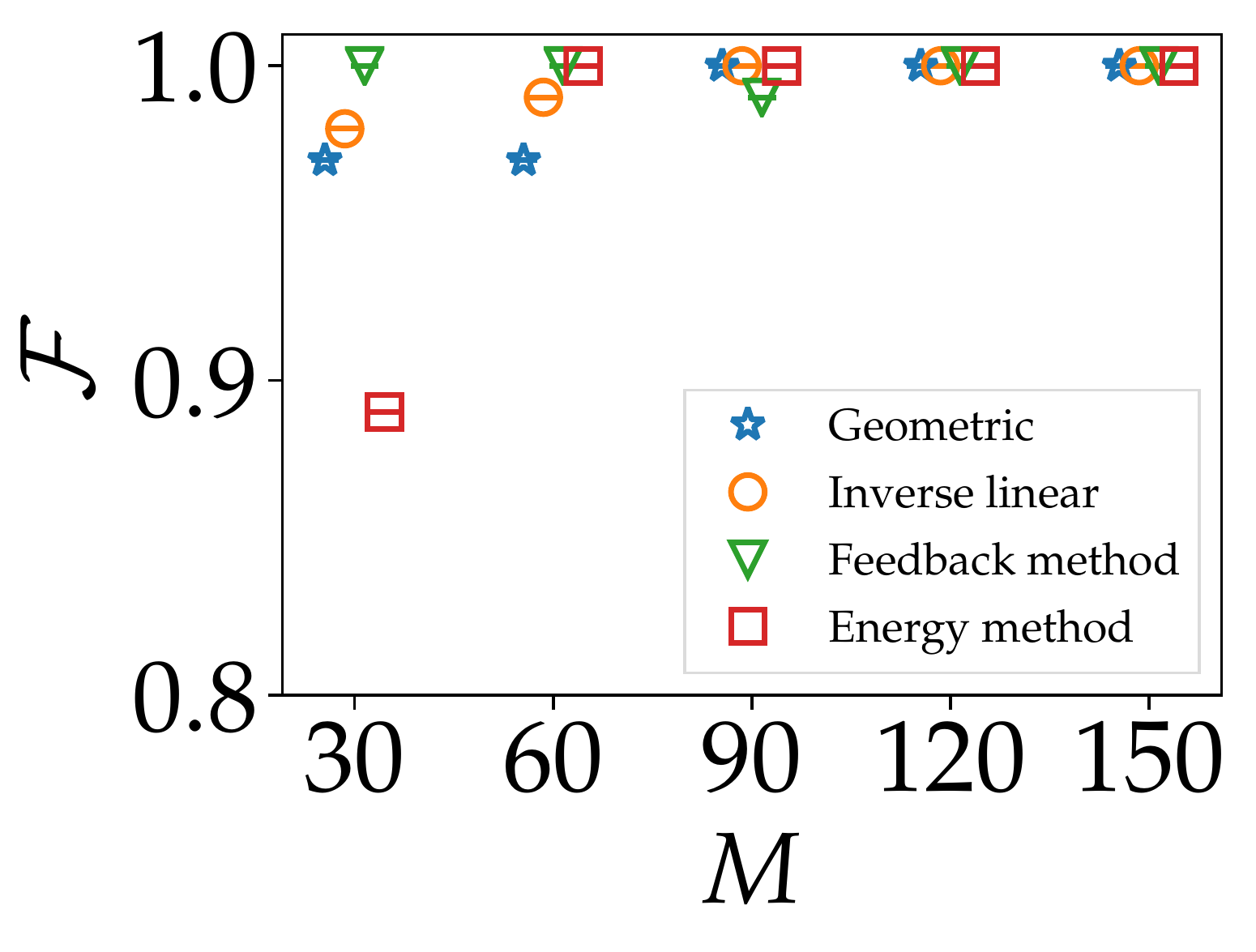}
\caption{\hspace{0.7cm}(a)}
\label{fig:SK_Wishart_frac_solved}
\end{subfigure} 
\begin{subfigure}[b]{0.49\linewidth}
\centering
\includegraphics[width=1\linewidth]{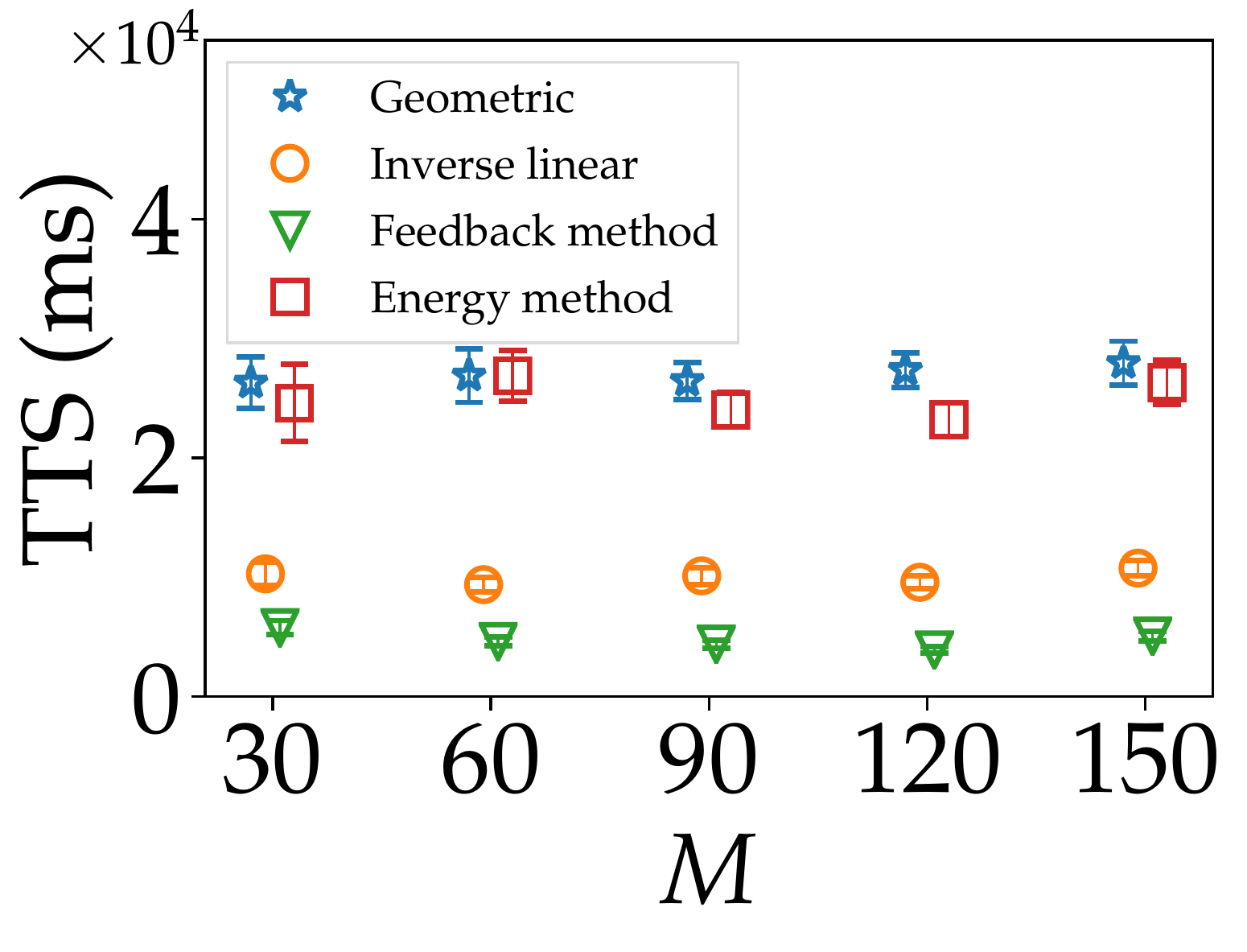}
\caption{\hspace{0.7cm}(b)}
\label{fig:SK_Wishart_tts_vs_M}
\end{subfigure}
\caption{(a) Fraction of problems solved, $\mathcal{F}$, and (b) median TTS versus the
 number of replicas, $M$, for various temperature setting methods and for Wishart 
 problems with $N=128$.}
\label{fig:num_temps_Wishart} 
\end{figure}

To investigate how the performance of the feedback-optimized method is affected as the number 
of replicas increases, the fraction of problems solved to optimality, $\mathcal{F}$, and the 
median TTS values are shown in Fig.~\ref{fig:num_temps_Wishart} for various numbers of replicas 
for a problem with $128$ variables. The feedback-optimized method solves all problem instances to optimality 
starting from $30$ replicas, whereas the other methods require at least $90$ replicas to solve 
all problem instances to optimality. The difference between the TTS of the feedback-optimized 
method and the other methods remains relatively constant for varying numbers of replicas. 

\section{Conclusion}
\label{sec:conclusion}
In this paper, we have compared the performance of four temperature setting 
techniques, the feedback-optimized method, the energy method, and two base methods 
(geometric and inverse-linear), all using random instances of spin-glass and Wishart problems. 

The feedback-optimized and the energy methods are dynamic approaches that rely on measuring 
statistical properties, such as local diffusivity and average energy, in respective terms, to 
iteratively adjust temperature values. The former method aims for a nonuniform replica-exchange 
probability by increasing the density of the temperature values around the simulation bottlenecks. The 
latter method, however, is designed to ensure constant probability across neighboring replicas. 

Our results support the superiority of the feedback-optimized method on Wishart problems 
that have first-order phase transitions at low temperatures. The replica-exchange probability has 
its highest density at the vicinity of the critical temperature ($T_c \simeq 0.25$). More specifically, the
feedback-optimized method results in a TTS speedup of a factor of two and five, compared to the 
inverse-linear method and to the energy and the geometric methods, respectively. 
On spin-glass problems with a continuous transition between the disordered and ordered 
phases at critical temperatures, the computational overhead of the two dynamic temperature setting 
methods, feedback-optimized and energy, does not yield any performance 
advantage over the base methods. In this case, the inverse-linear method that arranges more 
temperature values close to the lowest temperature outperforms the other methods. 

In summary, our work provides experimental evidence that the optimization problems, especially those 
with discontinuous transitions, are well suited to benefit from nonuniform and temperature-dependent 
acceptance probabilities for swap moves when using the PT algorithm.

\section*{Acknowledgement} 

The authors thank Gili Rosenberg and Salvatore Mandr\`a
for useful discussion, and Marko Bucyk for editorial help.  A portion of H.~G.~K.'s 
research is based on work supported in part by the
Office of the Director of National Intelligence (ODNI), Intelligence
Advanced Research Projects Activity (IARPA), via MIT Lincoln Laboratory
Air Force Contract No.~FA8721-05-C-0002.  The views and conclusions
contained herein are those of the authors and should not be interpreted
as necessarily representing the official policies or endorsements,
either expressed or implied, of ODNI, IARPA, or the U.S.~Government.
The U.S.~Government is authorized to reproduce and distribute reprints
for Governmental purpose notwithstanding any copyright annotation
thereon.

\bibliographystyle{apsrevtitle}
\bibliography{refs}

\begin{thebibliography}{31}
\expandafter\ifx\csname natexlab\endcsname\relax\def\natexlab#1{#1}\fi
\expandafter\ifx\csname bibnamefont\endcsname\relax
  \def\bibnamefont#1{#1}\fi
\expandafter\ifx\csname bibfnamefont\endcsname\relax
  \def\bibfnamefont#1{#1}\fi
\expandafter\ifx\csname citenamefont\endcsname\relax
  \def\citenamefont#1{#1}\fi
\expandafter\ifx\csname url\endcsname\relax
  \def\url#1{\texttt{#1}}\fi
\expandafter\ifx\csname urlprefix\endcsname\relax\def\urlprefix{URL }\fi
\providecommand{\bibinfo}[2]{#2}
\providecommand{\eprint}[2][]{\url{#2}}

\bibitem[{\citenamefont{Geyer}(1991)}]{Geyer91}
\bibinfo{author}{\bibfnamefont{C.}~\bibnamefont{Geyer}}, in
  \emph{\bibinfo{booktitle}{23rd Symposium on the Interface}}, edited by
  \bibinfo{editor}{\bibfnamefont{E.~M.} \bibnamefont{Keramidas}}
  (\bibinfo{publisher}{Interface Foundation}, \bibinfo{address}{Fairfax
  Station, VA}, \bibinfo{year}{1991}), p. \bibinfo{pages}{156}.

\bibitem[{\citenamefont{Swendsen and Wang}(1986)}]{Swendsen86}
\bibinfo{author}{\bibfnamefont{R.~H.} \bibnamefont{Swendsen}} \bibnamefont{and}
  \bibinfo{author}{\bibfnamefont{J.-S.} \bibnamefont{Wang}},
  \emph{\bibinfo{title}{Replica {M}onte {C}arlo simulation of spin-glasses}},
  \bibinfo{journal}{Phys. Rev. Lett.} \textbf{\bibinfo{volume}{57}},
  \bibinfo{pages}{2607} (\bibinfo{year}{1986}).

\bibitem[{\citenamefont{Hukushima and Nemoto}(1996)}]{Hukushima96}
\bibinfo{author}{\bibfnamefont{K.}~\bibnamefont{Hukushima}} \bibnamefont{and}
  \bibinfo{author}{\bibfnamefont{K.}~\bibnamefont{Nemoto}},
  \emph{\bibinfo{title}{Exchange {M}onte {C}arlo method and application to spin
  glass simulations}}, \bibinfo{journal}{J. Phys. Soc. Jpn.}
  \textbf{\bibinfo{volume}{65}}, \bibinfo{pages}{1604} (\bibinfo{year}{1996}).

\bibitem[{\citenamefont{Earl and Deem}(2005)}]{Earl05}
\bibinfo{author}{\bibfnamefont{D.~J.} \bibnamefont{Earl}} \bibnamefont{and}
  \bibinfo{author}{\bibfnamefont{M.~W.} \bibnamefont{Deem}},
  \emph{\bibinfo{title}{Parallel tempering: {T}heory, applications, and new
  perspectives}}, \bibinfo{journal}{Phys. Chem. Chem. Phys.}
  \textbf{\bibinfo{volume}{7}}, \bibinfo{pages}{3910} (\bibinfo{year}{2005}).

\bibitem[{\citenamefont{Katzgraber}(2009)}]{Katzgraber09e}
\bibinfo{author}{\bibfnamefont{H.~G.} \bibnamefont{Katzgraber}},
  \emph{\bibinfo{title}{Introduction to {M}onte {C}arlo methods}}
  (\bibinfo{year}{2009}), \bibinfo{note}{(arXiv:0905.1629)}.

\bibitem[{\citenamefont{Moreno et~al.}(2003)\citenamefont{Moreno, Katzgraber,
  and Hartmann}}]{moreno2003finding}
\bibinfo{author}{\bibfnamefont{J.}~\bibnamefont{Moreno}},
  \bibinfo{author}{\bibfnamefont{H.~G.} \bibnamefont{Katzgraber}},
  \bibnamefont{and} \bibinfo{author}{\bibfnamefont{A.~K.}
  \bibnamefont{Hartmann}}, \emph{\bibinfo{title}{Finding low-temperature states
  with parallel tempering, simulated annealing and simple monte carlo}},
  \bibinfo{journal}{International Journal of Modern Physics C}
  \textbf{\bibinfo{volume}{14}}, \bibinfo{pages}{285} (\bibinfo{year}{2003}).

\bibitem[{\citenamefont{Mandr{\`a} et~al.}(2016)\citenamefont{Mandr{\`a}, Zhu,
  Wang, Perdomo-Ortiz, and Katzgraber}}]{mandra2016strengths}
\bibinfo{author}{\bibfnamefont{S.}~\bibnamefont{Mandr{\`a}}},
  \bibinfo{author}{\bibfnamefont{Z.}~\bibnamefont{Zhu}},
  \bibinfo{author}{\bibfnamefont{W.}~\bibnamefont{Wang}},
  \bibinfo{author}{\bibfnamefont{A.}~\bibnamefont{Perdomo-Ortiz}},
  \bibnamefont{and} \bibinfo{author}{\bibfnamefont{H.~G.}
  \bibnamefont{Katzgraber}}, \emph{\bibinfo{title}{Strengths and weaknesses of
  weak-strong cluster problems: A detailed overview of state-of-the-art
  classical heuristics versus quantum approaches}}, \bibinfo{journal}{Physical
  Review A} \textbf{\bibinfo{volume}{94}}, \bibinfo{pages}{022337}
  (\bibinfo{year}{2016}).

\bibitem[{\citenamefont{Zhu et~al.}(2016)\citenamefont{Zhu, Fang, and
  Katzgraber}}]{Zhu16y}
\bibinfo{author}{\bibfnamefont{Z.}~\bibnamefont{Zhu}},
  \bibinfo{author}{\bibfnamefont{C.}~\bibnamefont{Fang}}, \bibnamefont{and}
  \bibinfo{author}{\bibfnamefont{H.~G.} \bibnamefont{Katzgraber}},
  \emph{\bibinfo{title}{{borealis - A generalized global update algorithm for
  Boolean optimization problems}}} (\bibinfo{year}{2016}),
  \bibinfo{note}{(arXiv:1605.09399)}.

\bibitem[{\citenamefont{{Rathore} et~al.}(2005)\citenamefont{{Rathore},
  {Chopra}, and {de Pablo}}}]{Rathore05}
\bibinfo{author}{\bibfnamefont{N.}~\bibnamefont{{Rathore}}},
  \bibinfo{author}{\bibfnamefont{M.}~\bibnamefont{{Chopra}}}, \bibnamefont{and}
  \bibinfo{author}{\bibfnamefont{J.~J.} \bibnamefont{{de Pablo}}},
  \emph{\bibinfo{title}{Optimal allocation of replicas in parallel tempering
  simulations}}, \bibinfo{journal}{J. Chem. Phys.}
  \textbf{\bibinfo{volume}{122}}, \bibinfo{pages}{024111}
  (\bibinfo{year}{2005}).

\bibitem[{\citenamefont{Hamze et~al.}(2010)\citenamefont{Hamze, Dickson, and
  Karimi}}]{Hamze10}
\bibinfo{author}{\bibfnamefont{F.}~\bibnamefont{Hamze}},
  \bibinfo{author}{\bibfnamefont{N.}~\bibnamefont{Dickson}}, \bibnamefont{and}
  \bibinfo{author}{\bibfnamefont{K.}~\bibnamefont{Karimi}},
  \emph{\bibinfo{title}{Robust parameter selection for parallel tempering}},
  \bibinfo{journal}{Int J Mod Phys C} \textbf{\bibinfo{volume}{21}},
  \bibinfo{pages}{603} (\bibinfo{year}{2010}).

\bibitem[{\citenamefont{Kofke}(2004)}]{Kofke04}
\bibinfo{author}{\bibfnamefont{D.~A.} \bibnamefont{Kofke}},
  \emph{\bibinfo{title}{{Comment on "The incomplete beta function law for
  parallel tempering sampling of classical canonical systems" [J. Chem. Phys.
  120, 4119 (2004)]}}}, \bibinfo{journal}{J. Chem. Phys.}
  \textbf{\bibinfo{volume}{121}}, \bibinfo{pages}{1167} (\bibinfo{year}{2004}).

\bibitem[{\citenamefont{Predescu et~al.}(2005)\citenamefont{Predescu, Predescu,
  and Ciobanu}}]{Predescu05}
\bibinfo{author}{\bibfnamefont{C.}~\bibnamefont{Predescu}},
  \bibinfo{author}{\bibfnamefont{M.}~\bibnamefont{Predescu}}, \bibnamefont{and}
  \bibinfo{author}{\bibfnamefont{C.}~\bibnamefont{Ciobanu}},
  \emph{\bibinfo{title}{On the efficiency of exchange in parallel tempering
  {M}onte {C}arlo simulations}}, \bibinfo{journal}{J. Phys. Chem. B}
  \textbf{\bibinfo{volume}{109}}, \bibinfo{pages}{4189} (\bibinfo{year}{2005}).

\bibitem[{\citenamefont{Fiore}(2008)}]{Fiore08}
\bibinfo{author}{\bibfnamefont{C.~E.} \bibnamefont{Fiore}},
  \emph{\bibinfo{title}{First-order phase transitions: {A} study through the
  parallel tempering method}}, \bibinfo{journal}{Phys. Rev. E}
  \textbf{\bibinfo{volume}{78}}, \bibinfo{pages}{041109}
  (\bibinfo{year}{2008}).

\bibitem[{\citenamefont{Machta}(2009)}]{Machta09}
\bibinfo{author}{\bibfnamefont{J.}~\bibnamefont{Machta}},
  \emph{\bibinfo{title}{Strengths and weaknesses of parallel tempering}},
  \bibinfo{journal}{Phys. Rev. E} \textbf{\bibinfo{volume}{80}},
  \bibinfo{pages}{056706} (\bibinfo{year}{2009}).

\bibitem[{\citenamefont{Machta and Ellis}(2011)}]{Machta11}
\bibinfo{author}{\bibfnamefont{J.}~\bibnamefont{Machta}} \bibnamefont{and}
  \bibinfo{author}{\bibfnamefont{R.}~\bibnamefont{Ellis}},
  \emph{\bibinfo{title}{Monte {C}arlo methods for rough free energy landscapes:
  {P}opulation annealing and parallel tempering}}, \bibinfo{journal}{J. Stat.
  Phys.} \textbf{\bibinfo{volume}{144}}, \bibinfo{pages}{541}
  (\bibinfo{year}{2011}).

\bibitem[{\citenamefont{Predescu et~al.}(2004)\citenamefont{Predescu, Predescu,
  and Ciobanu}}]{Predescu04}
\bibinfo{author}{\bibfnamefont{C.}~\bibnamefont{Predescu}},
  \bibinfo{author}{\bibfnamefont{M.}~\bibnamefont{Predescu}}, \bibnamefont{and}
  \bibinfo{author}{\bibfnamefont{C.~V.} \bibnamefont{Ciobanu}},
  \emph{\bibinfo{title}{The incomplete beta function law for parallel tempering
  sampling of classical canonical systems}}, \bibinfo{journal}{J. Chem. Phys}
  \textbf{\bibinfo{volume}{120 9}}, \bibinfo{pages}{4119}
  (\bibinfo{year}{2004}).

\bibitem[{\citenamefont{Hukushima}(1999)}]{Hukushima99}
\bibinfo{author}{\bibfnamefont{K.}~\bibnamefont{Hukushima}},
  \emph{\bibinfo{title}{Domain-wall free energy of spin-glass models: Numerical
  method and boundary conditions}}, \bibinfo{journal}{Phys. Rev. E}
  \textbf{\bibinfo{volume}{60}} (\bibinfo{year}{1999}).

\bibitem[{\citenamefont{Trebst et~al.}(2004)\citenamefont{Trebst, Huse, and
  Troyer}}]{Trebst04}
\bibinfo{author}{\bibfnamefont{S.}~\bibnamefont{Trebst}},
  \bibinfo{author}{\bibfnamefont{D.~A.} \bibnamefont{Huse}}, \bibnamefont{and}
  \bibinfo{author}{\bibfnamefont{M.}~\bibnamefont{Troyer}},
  \emph{\bibinfo{title}{{Optimizing the ensemble for equilibration in
  broad-histogram Monte Carlo simulations}}}, \bibinfo{journal}{Phys. Rev. E}
  \textbf{\bibinfo{volume}{70}} (\bibinfo{year}{2004}).

\bibitem[{\citenamefont{Katzgraber et~al.}(2006)\citenamefont{Katzgraber,
  Trebst, Huse, and Troyer}}]{Katzgraber06}
\bibinfo{author}{\bibfnamefont{H.~G.} \bibnamefont{Katzgraber}},
  \bibinfo{author}{\bibfnamefont{S.}~\bibnamefont{Trebst}},
  \bibinfo{author}{\bibfnamefont{D.~A.} \bibnamefont{Huse}}, \bibnamefont{and}
  \bibinfo{author}{\bibfnamefont{M.}~\bibnamefont{Troyer}},
  \emph{\bibinfo{title}{{{Feedback-optimized parallel tempering Monte
  Carlo}}}}, \bibinfo{journal}{J. Stat. Mech.}
  \textbf{\bibinfo{volume}{\normalfont{P03018}}} (\bibinfo{year}{2006}).

\bibitem[{\citenamefont{Hamze et~al.}(2019)\citenamefont{Hamze, Raymond,
  Pattison, and Katzgraber}}]{Hamze_in_prep}
\bibinfo{author}{\bibfnamefont{F.}~\bibnamefont{Hamze}},
  \bibinfo{author}{\bibfnamefont{J.}~\bibnamefont{Raymond}},
  \bibinfo{author}{\bibfnamefont{C.~A.} \bibnamefont{Pattison}},
  \bibnamefont{and} \bibinfo{author}{\bibfnamefont{H.~G.}
  \bibnamefont{Katzgraber}}, \emph{\bibinfo{title}{The {W}ishart planted
  ensemble: {A} tunably-rugged pairwise {I}sing model with a first-order phase
  transition}} (\bibinfo{year}{2019}),
  \bibinfo{note}{https://arxiv.org/abs/1906.00275}.

\bibitem[{\citenamefont{Wang et~al.}(2015)\citenamefont{Wang, Machta, and
  Katzgraber}}]{Wang15}
\bibinfo{author}{\bibfnamefont{W.}~\bibnamefont{Wang}},
  \bibinfo{author}{\bibfnamefont{J.}~\bibnamefont{Machta}}, \bibnamefont{and}
  \bibinfo{author}{\bibfnamefont{H.~G.} \bibnamefont{Katzgraber}},
  \emph{\bibinfo{title}{Comparing {M}onte {C}arlo methods for finding ground
  states of {I}sing spin glasses: {P}opulation annealing, simulated annealing,
  and parallel tempering}}, \bibinfo{journal}{Phys. Rev. E}
  \textbf{\bibinfo{volume}{92}}, \bibinfo{pages}{013303}
  (\bibinfo{year}{2015}).

\bibitem[{\citenamefont{Atkins and De~Paula}(2006)}]{Atkins06}
\bibinfo{author}{\bibfnamefont{P.}~\bibnamefont{Atkins}} \bibnamefont{and}
  \bibinfo{author}{\bibfnamefont{J.}~\bibnamefont{De~Paula}},
  \emph{\bibinfo{title}{Physical Chemistry for the Life Sciences}}
  (\bibinfo{publisher}{Oxford University Press}, \bibinfo{year}{2006}).

\bibitem[{\citenamefont{Steffen}(1990)}]{Steffen90}
\bibinfo{author}{\bibfnamefont{M.}~\bibnamefont{Steffen}},
  \emph{\bibinfo{title}{{{A simple method for monotonic interpolation in one
  dimension}}}}, \bibinfo{journal}{Astron. Astrophys.}
  \textbf{\bibinfo{volume}{239}} (\bibinfo{year}{1990}).

\bibitem[{\citenamefont{Pardella and Liers}(2008)}]{Pardella08}
\bibinfo{author}{\bibfnamefont{G.}~\bibnamefont{Pardella}} \bibnamefont{and}
  \bibinfo{author}{\bibfnamefont{F.}~\bibnamefont{Liers}},
  \emph{\bibinfo{title}{Exact ground states of large two-dimensional planar
  {I}sing spin glasses}}, \bibinfo{journal}{Phys. Rev. E}
  \textbf{\bibinfo{volume}{78}}, \bibinfo{pages}{056705}
  (\bibinfo{year}{2008}).

\bibitem[{\citenamefont{Liers and Pardella}(2010)}]{Liers10}
\bibinfo{author}{\bibfnamefont{F.}~\bibnamefont{Liers}} \bibnamefont{and}
  \bibinfo{author}{\bibfnamefont{G.}~\bibnamefont{Pardella}},
  \emph{\bibinfo{title}{Partitioning planar graphs: {A} fast combinatorial
  approach for max-cut}}, \bibinfo{journal}{Comput. Optim. Appl.}
  p.~\bibinfo{pages}{1} (\bibinfo{year}{2010}).

\bibitem[{\citenamefont{Elf et~al.}(2001)\citenamefont{Elf, Gutwenger,
  J{\"u}nger, and Rinaldi}}]{Juenger01}
\bibinfo{author}{\bibfnamefont{M.}~\bibnamefont{Elf}},
  \bibinfo{author}{\bibfnamefont{C.}~\bibnamefont{Gutwenger}},
  \bibinfo{author}{\bibfnamefont{M.}~\bibnamefont{J{\"u}nger}},
  \bibnamefont{and} \bibinfo{author}{\bibfnamefont{G.}~\bibnamefont{Rinaldi}},
  \emph{\bibinfo{title}{Lecture notes in computer science 2241}}, in
  \emph{\bibinfo{booktitle}{Computational Combinatorial Optimization}}, edited
  by \bibinfo{editor}{\bibfnamefont{M.}~\bibnamefont{J{\"u}nger}}
  \bibnamefont{and} \bibinfo{editor}{\bibfnamefont{D.}~\bibnamefont{Naddef}}
  (\bibinfo{publisher}{Springer Verlag}, \bibinfo{address}{Heidelberg},
  \bibinfo{year}{2001}), vol. \bibinfo{volume}{2241}.

\bibitem[{\citenamefont{Gr{\"o}tschel et~al.}(1987)\citenamefont{Gr{\"o}tschel,
  J{\"u}nger, and Reinelt}}]{Groetschel87}
\bibinfo{author}{\bibfnamefont{M.}~\bibnamefont{Gr{\"o}tschel}},
  \bibinfo{author}{\bibfnamefont{M.}~\bibnamefont{J{\"u}nger}},
  \bibnamefont{and} \bibinfo{author}{\bibfnamefont{G.}~\bibnamefont{Reinelt}},
  in \emph{\bibinfo{booktitle}{{Heidelberg colloquium on glassy dynamics}}}
  (\bibinfo{publisher}{Springer}, \bibinfo{year}{1987}), p.
  \bibinfo{pages}{325}.

\bibitem[{\citenamefont{Sherrington and Kirkpatrick}(1975)}]{Sherrington75}
\bibinfo{author}{\bibfnamefont{D.}~\bibnamefont{Sherrington}} \bibnamefont{and}
  \bibinfo{author}{\bibfnamefont{S.}~\bibnamefont{Kirkpatrick}},
  \emph{\bibinfo{title}{Solvable model of a spin glass}},
  \bibinfo{journal}{Phys. Rev. Lett.} \textbf{\bibinfo{volume}{35}},
  \bibinfo{pages}{1792} (\bibinfo{year}{1975}).

\bibitem[{\citenamefont{R{\o}nnow et~al.}(2014)\citenamefont{R{\o}nnow, Wang,
  Job, Boixo, Isakov, Wecker, Martinis, Lidar, and
  Troyer}}]{ronnow2014defining}
\bibinfo{author}{\bibfnamefont{T.~F.} \bibnamefont{R{\o}nnow}},
  \bibinfo{author}{\bibfnamefont{Z.}~\bibnamefont{Wang}},
  \bibinfo{author}{\bibfnamefont{J.}~\bibnamefont{Job}},
  \bibinfo{author}{\bibfnamefont{S.}~\bibnamefont{Boixo}},
  \bibinfo{author}{\bibfnamefont{S.~V.} \bibnamefont{Isakov}},
  \bibinfo{author}{\bibfnamefont{D.}~\bibnamefont{Wecker}},
  \bibinfo{author}{\bibfnamefont{J.~M.} \bibnamefont{Martinis}},
  \bibinfo{author}{\bibfnamefont{D.~A.} \bibnamefont{Lidar}}, \bibnamefont{and}
  \bibinfo{author}{\bibfnamefont{M.}~\bibnamefont{Troyer}},
  \emph{\bibinfo{title}{Defining and detecting quantum speedup}},
  \bibinfo{journal}{Science} \textbf{\bibinfo{volume}{345}},
  \bibinfo{pages}{420} (\bibinfo{year}{2014}).

\bibitem[{\citenamefont{Aramon et~al.}(2019)\citenamefont{Aramon, Rosenberg,
  Valiante, Miyazawa, Tamura, and Katzgraber}}]{Aramon19}
\bibinfo{author}{\bibfnamefont{M.}~\bibnamefont{Aramon}},
  \bibinfo{author}{\bibfnamefont{G.}~\bibnamefont{Rosenberg}},
  \bibinfo{author}{\bibfnamefont{E.}~\bibnamefont{Valiante}},
  \bibinfo{author}{\bibfnamefont{T.}~\bibnamefont{Miyazawa}},
  \bibinfo{author}{\bibfnamefont{H.}~\bibnamefont{Tamura}}, \bibnamefont{and}
  \bibinfo{author}{\bibfnamefont{H.~G.} \bibnamefont{Katzgraber}},
  \emph{\bibinfo{title}{Physics-inspired optimization for quadratic
  unconstrained problems using a digital annealer}},
  \bibinfo{journal}{Frontiers in Physics} \textbf{\bibinfo{volume}{7}},
  \bibinfo{pages}{48} (\bibinfo{year}{2019}).

\bibitem[{SG()}]{SG}
\bibinfo{note}{{http://informatik.uni-koeln.de/spinglass}}.

\end{thebibliography}

\end{document}